\documentclass[article]{JHEP3} 

\usepackage{epsfig,multicol,bbm}
\newcommand\fverb{\setbox\fverbbox=\hbox\bgroup\verb}
\newcommand\fverbdo{\egroup\medskip\noindent%
			\fbox{\unhbox\fverbbox}\ }
\newcommand\fverbit{\egroup\item[\fbox{\unhbox\fverbbox}]}
\newbox\fverbbox

\title{Quantum Tunneling, Blackbody Spectrum and Non-Logarithmic Entropy Correction for Lovelock Black Holes}

\author{Rabin Banerjee and Sujoy Kumar Modak\\
	S.N. Bose National Centre for Basic Sciences\\ Block-JD, Sector-III, Salt Lake City, Kolkata-700098, India \\
	E-mail: \email{rabin@bose.res.in\\}E-mail: \email{sujoy@bose.res.in}}

\abstract{We show, using the tunneling method, that Lovelock black holes Hawking radiate with a perfect blackbody spectrum. This is a new result. Within the semiclassical (WKB) approximation the temperature of the spectrum is given by the semiclassical Hawking temperature. Beyond the semiclassical approximation the thermal nature of the spectrum does not change but the temperature undergoes some higher order corrections. This is true for both black hole (event) and cosmological horizons. Using the first law of thermodynamics the black hole entropy is calculated. Specifically, the $D$-dimensional static, chargeless black hole solutions which are spherically symmetric and asymptotically flat, AdS or dS are considered. The interesting property of these black holes is that their semiclassical entropy does not obey the Bekenstein-Hawking area law. Furthermore, it is found that the leading correction to the semiclassical entropy for these black holes is not logarithmic and next to leading correction is also not inverse of horizon area. This is in contrast to the black holes in Einstein gravity. The modified result is due to the presence of Gauss-Bonnet term in the Lovelock Lagrangian. For the limit where the coupling constant of the Gauss-Bonnet term vanishes one recovers the known correctional terms as expected in Einstein gravity. Finally we relate the coefficient of the leading (non-logarithmic) correction with the trace anomaly of the stress tensor.}

\keywords{Black Holes, Classical Theories of Gravity}

\begin{document} 


\section{Introduction}

The study of black holes is one of the rich and modern subjects of physics, thanks to Einstein's theory of general relativity of gravitation. The special behavior of the black hole event horizon together with nontrivial aspects of space-time infinities have created a lot of interest in this subject even within classical gravity. After incorporating the quantum behavior of the fields propagating in a fixed curved background, Hawking (1975) \cite{Hawking} found that black holes were not really black as they were believed to be, rather they emit all kinds of particles with a perfect black body spectrum. His invention was inspired by the earlier works of Bekenstein \cite{Beken}. This remarkable discovery together with the first law of black hole mechanics \cite{Bardeen} established a strong connection with the first law of thermodynamics. Using this one can identify the black hole entropy as one quarter of it's horizon area. This is known as the celebrated ``Bekenstein-Hawking area law'', given by $S_{\textrm{BH}}=\frac{A}{4G\hbar}$. In a recent work \cite{Modak} we have made an analysis based on a purely thermodynamical viewpoint without using any analogy with the ``first law of black hole mechanics''. Considering the black hole entropy as a state function, a systematic calculation naturally led to the known area law. 

All known black holes in Einstein gravity in any dimension satisfy the area law. These black hole solutions are generally found by solving the Einstein field equation which does not contain any higher curvature term (since the Einstein-Hilbert action contains the Ricci scalar as the only scalar curvature). But one can always construct scalar curvatures by considering higher curvature terms and include those in the starting Lagrangian to generalize the Einstein-Hilbert action. Lovelock Lagrangian (\ref{1.1}) is one such natural generalization which is a sum over all such higher curvature terms with the Einstein-Hilbert and cosmological terms in the lowest order. The theory which includes only the first nontrivial higher curvature (Gauss-Bonnet) term to the other two lowest order terms in the Lagrangian is called the Einstein-Gauss-Bonnet (EGB) theory. In the equation of motion, obtained by varying the new action, one finds the Einstein equation modified by an extra part. Black hole solutions corresponding to this new equation of motion are called Lovelock black holes. In this paper we shall be dealing with the chargeless, static Lovelock black holes of the EGB theory \cite{Deser, Wheeler1, Wheeler2, Myers,Jacob,Cai,Odinstov2,Cai2}. The causal structures of Lovelock black hole spacetimes, which depend on the choice of parameters within it, sometime show some drastic differences with the black holes in Einstein gravity.  An interesting property of such Lovelock black holes is that they do not obey the usual semiclassical area law. The presence of the Gauss-Bonnet term in the action adds an extra additive term which can be written as a function of the horizon area. However, when the coupling constant of the Gauss-Bonnet term vanishes one recovers the standard area law again. 

Finding the corrections to the semiclassical Bekenstein-Hawking entropy, which can be very significant in Planck scale, has drawn a lot of interest these days. There exist several approaches to find these corrections. These are based on field theory \cite{Fursaev}, quantum geometry \cite{Partha}, statistical mechanics \cite{Das}, Cardy formula \cite{Carlip}, brick wall method \cite{Hooft} and tunneling method \cite{Modak, RBan, Majhibeyond, Modak1, Majhitrace}. Despite the diversity among the approaches they all agree on the logarithmic correction as the leading correction to the area law {\footnote{For an extensive list of papers on logarithmic correction {\it see} \cite{Page}}}. The only distinction occurs in the normalisation of the logarithmic term. While all these papers were confined to the black holes of Einstein gravity, surprisingly there is no such result which discusses possible quantum corrections to the semiclassical entropy for general Lovelock black holes. Furthermore, although it is usually considered that Lovelock black holes Hawking radiate, there is no analysis which directly yields the blackbody spectrum of radiation. The motivation of this paper is to address these issues.

Here we show, using the tunneling mechanism, that Lovelock black holes Hawking radiate both scalar particles and fermions with a perfect blackbody spectrum. The temperature of the radiation exactly matches with the semiclassical Hawking temperature of a black hole. Though tunneling method is widely used to calculate the semiclassical Hawking temperature of a black hole there was a glaring omission since it failed to yield directly the blackbody spectrum. However in a recent collaborative work involving one of us \cite{Majhiflux} this gap was filled and the blackbody distribution was reproduced in this mechanism, in the context of Einstein gravity. This work followed from a reformulation of the tunneling phenomena proposed in \cite{Majhiconnect}. In this paper we not only generalise these methods for the black hole horizon but also for the cosmological horizon corresponding to arbitrary $D$-dimensional spacetimes in Lovelock gravity. We then extend our method by going beyond the semiclassical approximation. It is shown that in presence of higher order corrections to the WKB ansatz the nature of the spectrum does not change. The spectrum remains purely thermal but the temperature receives higher order corrections to its semiclassical value. This is true for both the black hole (event) and cosmological horizons.

We study the thermodynamic properties of three different chargeless, static black holes in arbitrary $D$- dimensions which are asymptotically flat, AdS and dS. Using the first law of thermodynamics, the entropy is calculated. The lowest order contribution just reproduces the semiclassical expressions \cite{Myers,Cai,Cai2} for the entropy which are different from the area law found in Einstein gravity. Next, it is found that the leading correction to the semiclassical value is not logarithmic and next to leading correction is also not inverse of horizon area in general. The presence of the Gauss-Bonnet coupling, therefore, modifies the results of black holes in Einstein gravity in a nontrivial manner. In the limit where the coupling constant vanishes the familiar logarithmic and inverse horizon area terms reappear, as expected in Einstein gravity. The coefficient of the leading correction for each case is related with the trace anomaly of the stress tensor in $D$- dimensions.  

To put our analysis in a proper perspective we recall that Hawking radiation in the tunneling picture is usually described by two methods, namely, the radial null geodesic method \cite{Parikh} and the Hamilton-Jacobi method \cite{Paddy}. The second variant is also known as the method of complex path. A reformulation of these conventional approaches was necessary to yield the blackbody spectrum. Using this reformulation the modified spectrum for Lovelock black holes is calculated by going beyond the semiclassical approximation from which the corrected Hawking temperature also gets identified.  The use of the first law of thermodynamics then leads to the corrected entropy.

We organize this paper in the following way. In section \ref{sec:2} we introduce Lovelock black holes and discuss some of their properties. In section \ref{sec:3} we derive the blackbody spectrum of Hawking radiation for Lovelock black holes. Here we consider the radiation from both black hole (event) horizon and also from the cosmological horizon. The semiclassical Hawking temperatures are reproduced for different spacetimes. Section \ref{sec:4} is used for the calculation of the modified radiation spectrum by going beyond the semiclassical approximation. The corrected Hawking temperatures are found for different spacetimes. In section \ref{sec:5} the corrected entropy is calculated for each spacetime. The standard semiclassical expressions for the temperature and entropy are also reproduced in these sections. The relation of the coefficient of the leading correction to the trace anomaly is established in section \ref{sec:6}. Lastly we give our concluding remarks in section \ref{sec:7}. We give a detailed description of our notations and definitions in three appendices (\ref{sec:A}), (\ref{sec:B}) and (\ref{sec:C}).

\section{Lovelock black holes}{\label{sec:2}}
Lovelock gravity is the most natural generalization of Einstein gravity in higher dimensions. Lovelock Lagrangian is the sum of dimensionally extended Euler densities, given by \cite{Lovelock1, Lovelock2}
\begin{equation}
{\cal L}=\sqrt{-g}\displaystyle\sum_{m=0}^{m=n} {c_{m} L_{m}},
\label{1.1}
\end{equation}
where,
\begin{equation}
L_m= 2^{-m}\delta^{\mu_1 \nu_1.....\mu_m \nu_m}_{\alpha_1\beta_1.....\alpha_m\beta_m}R^{\alpha_1\beta_1}_{~~~~~{\mu_1\nu_1}}.....R^{\alpha_m\beta_m}_{~~~~~{\mu_m\nu_m}}.
\label{1.2}
\end{equation}
The generalized Kronecker $\delta$-function is totally antisymmetric in both sets of indices,
\begin{equation}
\delta^{\mu_1 \nu_1.....\mu_m \nu_m}_{\alpha_1\beta_1.....\alpha_m\beta_m}=\frac{1}{m!}\delta^{\mu_1 \nu_1.....\mu_m \nu_m}_{[\alpha_1\beta_1.....\alpha_m\beta_m]}.
\label{1.3}
\end{equation}  
The coupling constants $c_m$ have dimensions of $[{\textrm {length}}]^{2m-D}$. Let us first introduce the different terms in the Lovelock Lagrangian (\ref{1.1}). The first term ($L_0$) is usually set to unity and therefore $c_0$ represents the cosmological constant ($\Lambda$). An explicit computation yields $L_1=R$ which represents the standard Einstein-Hilbert term while $L_2=(R^2+R_{\alpha\beta\mu\nu}R^{\alpha\beta\mu\nu}-4R_{\mu\nu}R^{\mu\nu})$ is the Gauss-Bonnet term. The Gauss-Bonnet term is a topological invariant in four dimensions and has nontrivial effects for $D>4$. The theory involving only these three terms is known as Einstein-Gauss-Bonnet (EGB) theory. The interesting property of the Lovelock Lagrangian is that despite the presence of the Riemann tensor in the Lovelock Lagrangian, the equation of motion does not contain any derivative of Riemann tensor (i.e. only upto second derivative of the metric tensor). In this sense Lovelock gravity is very similar to Einstein gravity since in Einstein equation one has terms only upto second derivative of the metric tensor. Also, this theory is known to be free of ghosts when expanding around a flat space \cite{Deser} thereby evading any problems with unitarity.

In the present paper we shall deal with all the chargeless static black hole solutions of EGB theory. The black hole solutions characterized by the mass parameter ($M$) only can be categorized in three different classes. One of them is the asymptotically flat solution excluding the cosmological constant, whereas, the other two are de Sitter and anti-de Sitter solutions depending on the sign of the cosmological constant when it is included. In the next section we give a collective information and comparison among the behaviour and thermodynamic properties of these black holes. These results are already present in the literature but we shall bring them together to construct the platform for the remaining work.

\subsection{Spherically symmetric, asymptotically flat black holes}
Considering only the Einstein-Hilbert and the Gauss-Bonnet term in (\ref{1.1}) we find the simplest higher-derivative Lovelock action which is given by 
\begin{equation}
I=\frac{1}{16\pi G}\int{d^Dx\sqrt{-g}\left(R+\frac{\lambda}{2}(R^2+R_{\alpha\beta\mu\nu}R^{\alpha\beta\mu\nu}-4R_{\mu\nu}R^{\mu\nu})\right)}
\label{1.4}
\end{equation}
where $\lambda$ (=$2c_2$ in (\ref{1.1})) is the coupling constant of the Gauss-Bonnet term having dimension of $({\textrm{length}})^2$ and is positive in the heterotic string theory \cite{Deser}.
The spherically symmetric vacuum solution of the equation of motion corresponding to the above action is given by \cite{Deser, Wheeler1, Wheeler2, Myers}
\begin{equation}
ds^2=-f(r) dt^2+\frac{1}{f(r)} dr^2+ r^2d\Omega_{D-2}^2, 
\label{1.5}
\end{equation}
with 
\begin{equation}
f(r)=1+\frac{r^2}{\bar\lambda}\left[1+\epsilon\left(1+\frac{2\omega\bar\lambda}{r^{D-1}}\right)^{\frac{1}{2}}\right],
\label{1.6}
\end{equation}
where $\bar\lambda=\lambda(D-3)(D-4)$. The constant $\epsilon$ has the value of $\pm1$. Only $\epsilon=-1$ corresponds to an asymptotically flat metric. Here $\omega$ appears as an integration constant and for $\epsilon=-1$ (i.e. for the asymptotically flat case) it  is related to the mass of the black hole as \cite{Myers}
\begin{eqnarray}
M=\frac{(D-2)A_{D-2}}{16\pi G}\omega,
\label{1.6a}
\end{eqnarray}
where $A_{D-2}$ is the $(D-2)$ dimensional volume element of the hypersurface parametrized by the angular variables in (\ref{1.5}). The positions of the horizons are found by setting $g_{tt}(r=r_h)=g^{rr}(r=r_h)=0$, which leads to the polynomial 
\begin{equation}
r_h^{D-3}+\frac{\bar\lambda}{2}r_h^{D-5}-\omega=0.
\label{1.7}
\end{equation}
Thus the real roots of (\ref{1.7}) give the positions of horizons and the largest root of them is the event horizon while the others are inner Cauchy horizons. Putting the value of the integration constant $\omega$ from (\ref{1.7}) into (\ref{1.6a}), the mass of this black hole can be expressed in terms of the radius of the event horizon as
\begin{eqnarray}
M=\frac{(D-2)A_{D-2}}{16\pi G}\left(r_h^{(D-3)}+\frac{\bar\lambda}{2}r_h^{(D-5)}\right)
\label{1.7a}
\end{eqnarray} 

The thermodynamical entities like Hawking temperature and semiclassical entropy for the above black hole solution in arbitrary `$D$'-dimensions have been calculated in \cite{Myers,Jacob}, given by
\begin{eqnarray}
T_{\textrm H}=\frac{\hbar (D-3)}{4\pi r_h}\left[\frac{r_h^2+\frac{D-5}{D-3}\frac{\bar\lambda}{2}}{r_h^2+\bar\lambda}\right]
\label{1.8}
\end{eqnarray}
and 
\begin{eqnarray}
S=\frac{A_{D-2}r_h^{(D-2)}}{4G\hbar}\left(1+\frac{D-2}{D-4}\frac{\bar\lambda}{r_h^2}\right)
\label{1.9}
\end{eqnarray}
respectively. Their derivation \cite{Myers} follows the original arguments of Hawking based on avoiding the conical singularity. One can see from (\ref{1.9}) that these black holes do not obey the Bekenstein-Hawking area law which states that entropy is one quarter of its horizon area. The presence of the Gauss-Bonnet term to the standard Einstein-Hilbert term modifies the area law upto an additive term which can itself be written as some function of horizon area for any dimension $D>4$. However in $D=4$ the additional term multiplied by $\bar\lambda$ (see (\ref{1.9})) is only an additive constant given by $\frac{2\pi\lambda}{G\hbar}$.

It is worthwhile to mention that the semiclassical result for black hole entropy for Lovelock black holes has also been derived using Wald's entropy formula \cite{Wald}, given by
\begin{eqnarray}
S=-2\pi\int_{\Sigma}\frac{\delta{\tilde{\cal L}}}{\delta R_{abcd}}\epsilon_{ab}\epsilon_{cd},
\label{waldentropy}
\end{eqnarray}
where the diffeomorphism-invariant Lagrangian ($\tilde{\cal L}$) is constructed by using any combination of curvature invariants. Here $\epsilon_{ab}$ is the binormal to the bifurcation (D-2) surface $\Sigma$ and the integral is taken with respect to the natural, induced volume element on $\Sigma$. It is found that the semiclassical entropy contains an extra term other than $\frac{A}{4G\hbar}$. For the Gauss-Bonnet black holes the semiclassical entropy has been calculated in \cite{Jacob} which also shows a modification to the usual area law. The additional term is the Euler constant of the cross section of the horizon. In four dimensions, this constant is fixed for all stationary black holes since the horzon topology is $S^2\times R$. For $D>4$ the horizon topology is not unique and this topological contribution to the entropy has more importance.

\subsection{Topological Gauss-Bonnet AdS black holes}
The action consisting of the Einstein-Hilbert plus the Gauss-Bonnet terms with a negative cosmological constant $\Lambda=-(D-1)(D-2)/2l^2$ in $D$ dimensions, is given by
\begin{equation}
I=\frac{1}{16\pi G}\int{d^Dx\sqrt{-g}\left(R+\alpha(R^2+R_{\alpha\beta\mu\nu}R^{\alpha\beta\mu\nu}-4R_{\mu\nu}R^{\mu\nu})+\frac{(D-1)(D-2)}{l^2}\right)}
\label{1.9a}
\end{equation}
Considering this action Cai \cite{Cai} found the AdS black hole solution in EGB theory. This is given by the metric which is not necessarily spherically symmetric,
\begin{eqnarray}
ds^2=-F(r)dt^2+\frac{1}{F(r)}dr^2+r^2h_{ij}dx^idx^j,
\label{1.10}
\end{eqnarray}
with
\begin{eqnarray}
F(r)=k+\frac{r^2}{2\tilde\alpha}\left(1\mp\sqrt{1+\frac{64\pi G\tilde\alpha M}{(D-2)\Sigma_{k}r^{D-1}}-\frac{4\tilde\alpha}{l^2}}\right)
\label{1.11}
\end{eqnarray}
where $\tilde\alpha=\alpha(D-3)(D-4)$. The possible values that $k$ can take are 1, 0 and $-1$ for which the $(D-2)$ dimensional metric in (\ref{1.9a}) yields spherical, plannar and hyperbolic symmetry respectively. Here $\Sigma_k$ is the volume element of the $(D-2)$ dimensional hypersurface parametrized by the angular variables in (\ref{1.10}) and for the special case $k=1$, i.e. for the spherically symmetric case, it exactly matches with $A_{D-2}$ in (\ref{1.7a}). The largest root of $F(r)=0$ defines the event horizon for this case and the mass of the black hole can be written in terms of the horizon radius as
\begin{equation}
M=\frac{(D-2)\Sigma_kr_h^{(D-3)}}{16\pi G}\left(k+\frac{\tilde\alpha k^2}{r_h^2}+\frac{r_h^2}{l^2}\right).
\label{1.11a}
\end{equation}
In \cite{Cai} the Hawking temperature was calculated by the requirement of the absence of conical singularity at the horizon in the Euclidean sector of the black hole solution. The result is, 
\begin{eqnarray}
T_{\textrm{H}}^{\textrm{(AdS)}}=\frac{\hbar[(D-1)r_h^4+(D-3)kl^2r_h^2+(D-5)\tilde\alpha k^2l^2]}{4\pi l^2r_h(r_h^2+2\tilde\alpha k)}
\label{1.12}
\end{eqnarray}
The semiclassical entropy was calculated by assuming that entropy of the black hole must obey the first law of thermodynamics, leading to,
\begin{eqnarray}
S^{\textrm{(AdS)}}=\frac{\Sigma_kr_h^{(D-2)}}{4G\hbar}\left(1+\frac{(D-2)}{(D-4)}\frac{2\tilde\alpha k}{r_h^2}\right).
\label{1.13}
\end{eqnarray}
 The same expression for the semiclassical entropy was also found in \cite{Olea3} by using the regularization of the Euclidean action in a background independent method.

Let us now compare the two kinds of black hole solutions mentioned in the last two sections. Comparing the two actions (\ref{1.4}) and (\ref{1.9a}) we see that for the parameter values $\alpha=\frac{\lambda}{2}$, $k=1$ and $\frac{1}{l^2}=0$, equation (\ref{1.9a}) exactly matches with (\ref{1.4}). Therefore in this limiting case there is no difference in the thermodynamical behaviors between these two solutions. This is due Birkhoff's theorem which is well applicable for these one parameter spherically symmetric black hole solutions in Lovelock gravity \cite{Robin}.

\subsection{Gauss-Bonnet dS black holes}
If one includes a positive cosmological constant to the Einstein-Hilbert and Gauss-Bonnet terms, given by $\Lambda=(D-1)(D-2)/2l^2$, in arbitrary $D$ dimensions, the action becomes 
\begin{eqnarray}
I=\frac{1}{16\pi G}\int{d^Dx\sqrt{-g}\left(R+\alpha(R^2+R_{\alpha\beta\mu\nu}R^{\alpha\beta\mu\nu}-4R_{\mu\nu}R^{\mu\nu})-\frac{(D-1)(D-2)}{l^2}\right)}.
\label{1.14}
\end{eqnarray}
In \cite{Cai2} Cai and Guo discovered the spherically symmetric dS black hole solution corresponding to this action, given by, 
\begin{eqnarray}
ds^2=-\tilde F(r) dt^2+\frac{dr^2}{\tilde F(r)}+r^2d\Omega_{(D-2)}^2,
\label{1.15}
\end{eqnarray}
with
\begin{eqnarray}
\tilde F(r)=1+\frac{r^2}{2\tilde\alpha}\left(1\mp\sqrt{1+\frac{64\pi G\tilde\alpha M}{(D-2)A_{D-2}r^{D-1}}+\frac{4\tilde\alpha}{l^2}}\right).
\label{1.16}
\end{eqnarray}
Here $\tilde\alpha=\alpha(D-3)(D-4)$ and $M$ is an integration constant that matches with the AD mass \cite{Abbot} of the solution.

The solution for $M=0$ is given by the metric coefficient 
\begin{equation}
\tilde F(r)=1+\frac{r^2}{2\tilde\alpha}{\left(1\mp\sqrt{1+\frac{4\tilde\alpha}{l^2}}\right)}.
\label{1.17}
\end{equation}
Depending on the effective curvature radius $\frac{1}{l_{\textrm{eff}}^2}=-\frac{1}{2\tilde\alpha}\left(1\mp\sqrt{1+\frac{4\tilde\alpha}{l^2}}\right)$, one has both asymptotically dS and AdS cases for the massless solution (even if in the action only the positive cosmological constant was considered). When $l_{\textrm{eff}}^2$ is positive (negative), one has asymptotically dS (AdS) solution. For $\tilde\alpha> 0$ the solution is asymptotically dS for the branch with ``$-$'' sign in (\ref{1.17}) and for ``$+$'' sign it is asymptotically AdS. On the other hand for $\tilde\alpha< 0$ one has asymptotically dS solution for both branches subjected to the condition $\frac{\tilde\alpha}{l^2}\geq-1/4$. However, similar to the other two cases, here also the graviton is a ghost on the background for the ``$+$'' branch of (\ref{1.16}) and thus has less physical importance. The branch with ``$-$'' sign, to be considered in this paper, is always asymptotically dS for any sign of $\tilde\alpha$ (however for the ``$-$'' sign, $\frac{\tilde\alpha}{l^2}\geq-1/4$ has to be satisfied). 

For the massless Gauss-Bonnet dS black holes there is only one cosmological horizon at $r_c=l_{\textrm{eff}}$ and as the mass ($M$) increases a black hole horizon ($r_h$) arises while the cosmological horizon ($r_c$) shrinks. The black hole horizon always remains inside the cosmological horizon. For the black hole horizon the gravitational mass is given by the AD mass which is always positive and for the cosmological horizon it is given by the BBM mass \cite{BBM} which is just negative of the AD mass. With these definitions, the first law of thermodynamics is valid for both horizons when they are considered as different thermodynamical systems.

\subsubsection{Thermodynamics of the black hole horizon}
The position of the black hole horizon is given by the second largest root of $\tilde F(r=r_h=0)$. The gravitational mass, represented by the AD mass, is given in terms of the black hole horizon as
\begin{equation}
M=\frac{(D-2)A_{D-2}r_h^{D-3}}{16\pi G}(1+\frac{\tilde\alpha}{r_h^2}-\frac{r_h^2}{l^2})
\label{1.17a}
\end{equation}
The requirement of the absence of conical singularity at the black hole horizon in the Euclidean sector of the above black hole gives the semiclassical Hawking temperature
\cite{Cai2}
\begin{eqnarray}
T_{\textrm H}^{\textrm{(dS)}}=\frac{\hbar[(D-5)\tilde\alpha l^2-(D-1)r_h^4+(D-3)l^2r_h^2]}{4\pi l^2r_h(r_h^2+2\tilde\alpha)}
\label{1.18}
\end{eqnarray}
whereas, the semiclassical entropy is given by
\begin{eqnarray}
S^{\textrm{(dS)}}=\int\frac{dM}{T^{(\textrm{dS})}}=\frac{A_{D-2}r_h^{(D-2)}}{4G\hbar}\left(1+\frac{(D-2)}{(D-4)}\frac{2\tilde\alpha}{r_h^2}\right).
\label{1.19}
\end{eqnarray}

\subsubsection{Thermodynamics of the cosmological horizon}
The temperature of the cosmological horizon, calculated by the same process as stated earlier for the black hole horizon, is given by \cite{Cai2}
\begin{eqnarray}
T_{\textrm{c}}^{\textrm{(dS)}}=\frac{\hbar[-(D-5)\tilde\alpha l^2+(D-1)r_c^4-(D-3)l^2r_c^2]}{4\pi l^2r_c(r_c^2+2\tilde\alpha)}.
\label{1.20}
\end{eqnarray}
There is an overall sign difference between this and temperature of the black hole horizon. Since the total energy associated with the cosmological horizon (BBM energy) is just negative of AD energy as given in (\ref{1.17a}), the first law $dM=TdS$ is well applicable here. The semiclassical entropy associated with the cosmological horizon is now given by  

\begin{eqnarray}
S_{\textrm{c}}^{\textrm{(dS)}}=\frac{A_{D-2}r_c^{(D-2)}}{4G\hbar}\left(1+\frac{(D-2)}{(D-4)}\frac{2\tilde\alpha}{r_c^2}\right).
\label{1.21}
\end{eqnarray}


\section{Tunneling mechanism and blackbody spectrum}{\label{sec:3}}
Tunneling mechanism is widely used to study Hawking radiation. However the different variants \cite{Parikh,Paddy} of the tunneling mechanism only yield the temperature and not the spectrum. This was highlighted in a recent work involving one of us \cite{Majhiflux}. There it was shown for the first time that the tunneling formalism also provides the perfect blackbody spectrum of the radiation with the temperature given by the Hawking temperature of a black hole. Later this method was used successfully for other non-spherically symmetric spacetimes in Einstein gravity \cite{Umetsu}. But these works were strictly confined to the Einstein gravity only. Now we shall generalize the method to find the blackbody spectrum and Hawking temperature for the Lovelock black holes which were introduced in the last section. This will make the tunneling method more reliable in the context of Lovelock gravity.

Let us first consider a spherically symmetric, static spacetime represented by the metric
\begin{eqnarray}
ds^2=-g(r)dt^2+\frac{1}{g(r)}dr^2+r^2d\Omega_{D-2}^2
\label{metric}
\end{eqnarray}
The positions of the horizons can be found by solving $g(r=r_h)=0$. For the asymptotically flat and AdS black holes there is only one event horizon given by the largest root of this equation. For the dS case the largest root gives the position of cosmic horizon and the second largest root is the black hole event horizon.

We start our analysis by considering the massless scalar particle governed by the Klein-Gordon equation with the spacetime metric given by (\ref{metric}),     
\begin{equation}
-\frac{\hbar^2}{\sqrt{-g}}{\partial_\mu[g^{\mu\nu}\sqrt{-g}\partial_{\nu}]\Phi}=0.
\label{2.1}
\end{equation} 
Since in our analysis we shall be dealing with the radial trajectory it is enough to consider the $r-t$ sector of the metric (\ref{metric}) to solve (\ref{2.1}). This equation cannot be solved exactly, therefore we choose the standard (WKB) ansatz for $\Phi$ as
\begin{eqnarray} 
\Phi(r,t)=\exp[-\frac{i}{\hbar}{{\cal S}(r.t)}], 
\label{2.2}
\end{eqnarray}
where the action is expanded in powers of $\hbar$;
\begin{eqnarray}
{\cal S}(r,t)={\cal S}_{0}(r,t)+\displaystyle\sum_{i=1}^{\infty}\hbar^{i}{\cal S}_{i}(r,t).
\label{action}
\end{eqnarray} 
Now plugging (\ref{2.2}) into (\ref{2.1}) we get
\begin{eqnarray}
&&\frac{i}{{g(r)}}\Big(\frac{\partial {\cal S}}{\partial t}\Big)^2 - ig(r)\Big(\frac{\partial {\cal S}}{\partial r}\Big)^2 - \frac{\hbar}{g(r)}\frac{\partial^2 {\cal S}}{\partial t^2} + \hbar g(r)\frac{\partial^2 {\cal S}}{\partial r^2}
\nonumber
\\
&&+ \hbar{\frac{\partial g(r)}{\partial r}}\frac{\partial {\cal S}}{\partial r}=0.
\label{2.3}
\end{eqnarray}
Taking the semiclassical limit ($\hbar\rightarrow 0$), we obtain the first order partial differential equation,
\begin{eqnarray}
\frac{\partial {\cal S}_0}{\partial t}=\pm {g(r)}\frac{\partial {\cal S}_0}{\partial r}.
\label{semi}
\end{eqnarray}
This is nothing but the semiclassical Hamilton-Jacobi equation. We choose the semiclassical action for a scalar field moving under the background metric (\ref{metric}) in the same spirit as usually done in the semiclassical Hamilton-Jacobi theory. Looking at the time translational symmetry of the spacetime (\ref{metric}) we take the form of the semiclassical action as
\begin{eqnarray}
{\cal S}_{0}(r,t)=\Omega t+{\cal S}_{0}(r),
\label{actionform}
\end{eqnarray} 
where $\Omega$ is the conserved quantity corresponding to the time translational Killing vector field. It is identified as the effective energy experienced by the particle. Now substituting this in (\ref{semi}) one can easily find
\begin{eqnarray}
{\cal S}_{0}(r)=\pm\Omega\int\frac{dr}{g(r)},
\label{s0}
\end{eqnarray}
Using (\ref{s0}) in (\ref{actionform}) we finally find the semiclassical action as
\begin{eqnarray}
{\cal S}_{0}(r,t)=\Omega(t\pm\int\frac{dr}{g(r)}).
\label{action2}
\end{eqnarray} 
Now we have the solution for the scalar field (\ref{2.2}), 
\begin{eqnarray} 
\Phi(r,t) &=& e^{[-{\frac{i}{\hbar}}\Omega(t\pm\int\frac{dr}{g(r)})]}\nonumber\\
          &=& e^{-\frac{i}{\hbar}\Omega(t\pm r^*)} 
\label{soln}
\end{eqnarray}
expressed in terms of the tortoise coordinate $r^*=\int\frac{dr}{g(r)}$. We shall require this solution in our analysis to find the radiation spectrum for Lovelock black holes.


\subsection{Blackbody spectrum for the black hole (event) horizon}
The first step in the analysis is to find a coordinate system which is regular at the event horizon. We do not need to know the global behaviour of the spacetime. If we can find a coordinate system in which the metric (\ref{metric}) is defined both inside and outside the event horizon the purpose is solved. In such a coordinate system we can readily connect the right and left moving modes defined inside and outside the event horizon {\footnote{We use the subscript ``$in$'' for modes inside the event horizon. Since the observer stays outside the event horizon we use the subscript ``$obs$'' for modes outside the event horizon.}}. In appendix (\ref{sec:A}) a Kruskal-like extension of (\ref{metric}) is done by concentrating on the behavior at the black hole event horizon only. Earlier, a similar formulation in the context of Lovelock black holes was done in \cite{Wheeler1}. In appendix (\ref{sec:C}) we show how to identify the left and right moving modes inside and outside the event horizon. 

First, it can be noted that the set of coordinate transformations, given in (\ref{transformation}) and (\ref{transformation2}) are connected by the following transformations in ($t,r^{*}$) coordinate
{\begin{eqnarray}
t_{in} &=& t_{obs}-\frac{i\pi}{2\kappa}\nonumber\\
r^*_{in} &=&  r^*_{obs}+\frac{i\pi}{2\kappa}
\label{connect1}
\end{eqnarray}
In the null coordinates ($u,v$) (\ref{null},\ref{null2})these two relations are recast as
{\begin{eqnarray}
u_{in} &=& u_{obs}-\frac{i\pi}{\kappa}\nonumber\\
v_{in} &=& v_{obs}.
\label{connect2}
\end{eqnarray}
Following \cite{Majhiflux}, we consider a situation where $n$ number of non-interacting virtual pairs are created inside the black hole horizon. The left and right moving modes inside the black hole horizon, found in (\ref{soln}), are then given by (\ref{conv3}) and (\ref{conv4}) respectively. Here the null coordinates are defined in (\ref{null2}) with $r^*$ given by (\ref{tortoise3}). Likewise the left and right moving modes outside the event horizon are given by (\ref{conv1}) and (\ref{conv2}) respectively. Now it is easy to see that due to the transformation (\ref{connect2}) the inside and the outside modes are connected by,
\begin{eqnarray}
\Phi_{in}^{(R)} &=& {e^{-\frac{\pi\Omega}{\hbar\kappa}}}\Phi_{{obs}}^{{(R)}} \nonumber\\
\Phi_{in}^{(L)} &=& {\Phi_{{obs}}^{{(L)}}}
\label{connect3}
\end{eqnarray}

Any physical state corresponding to $n$ number of virtual pairs inside the black hole event horizon, when observed by an observer outside the horizon, is given by,
\begin{eqnarray}
|\Psi\rangle = N \displaystyle\sum_n |n^{(L)}_{\textrm{in}}\rangle\otimes|n^{(R)}_{\textrm{in}}\rangle
 = N \displaystyle\sum_n e^{-\frac{\pi n\Omega}{\hbar\kappa}}|n^{(L)}_{\textrm{obs}}\rangle\otimes|n^{(R)}_{\textrm{obs}}\rangle,
\label{state}
\end{eqnarray}
where $N$ is a normalisation constant. Here we have used the transformations (\ref{connect3}). Now using the normalization condition $\langle\Psi|\Psi\rangle=1$ and considering $n=0,1,2...$ for bosons or $n=0,1$ for fermions, one obtains
\begin{eqnarray}
N_{\textrm{(boson)}} &=& \Big(1-e^{-\frac{2\pi\Omega}{\hbar\kappa}}\Big)^{\frac{1}{2}}\label{norm1}\\
N_{\textrm{(fermion)}} &=& \Big(1+e^{-\frac{2\pi\Omega}{\hbar\kappa}}\Big)^{-\frac{1}{2}}
\label{norm2}
\end{eqnarray}
Therefore the physical states for them, viewed by an external observer, are given by
\begin{eqnarray}
|\Psi\rangle_{(\textrm{boson})}= \Big(1-e^{-\frac{2\pi\Omega}{\hbar\kappa}}\Big)^{\frac{1}{2}} \displaystyle\sum_n e^{-\frac{\pi n\Omega}{\hbar\kappa}}|n^{(L)}_{\textrm{obs}}\rangle\otimes|n^{(R)}_{\textrm{obs}}\rangle
\label{boson}
\\
|\Psi\rangle_{(\textrm{fermion})}= \Big(1+e^{-\frac{2\pi\Omega}{\hbar\kappa}}\Big)^{-\frac{1}{2}} \displaystyle\sum_n e^{-\frac{\pi n\Omega}{\hbar\kappa}}|n^{(L)}_{\textrm{obs}}\rangle\otimes|n^{(R)}_{\textrm{obs}}\rangle.
\label{fermion}
\end{eqnarray}
The density matrix operator for the bosons can be constructed as
\begin{eqnarray}
{\hat\rho}_{(\textrm{boson})}&=&|\Psi\rangle_{(\textrm{boson})}\langle\Psi|_{(\textrm{boson})}
\nonumber
\\
&=&\Big(1-e^{-\frac{2\pi\Omega}{\hbar\kappa}}\Big) \displaystyle\sum_{n,m} e^{-\frac{\pi (n+m)\Omega}{\hbar\kappa}} |n^{(L)}_{\textrm{obs}}\rangle\otimes|n^{(R)}_{\textrm{obs}}\rangle\langle m^{(R)}_{\textrm{obs}}|\otimes\langle m^{(L)}_{\textrm{obs}}|
\label{density}
\end{eqnarray}
Since the left going modes inside the horizon do not reach the outer observer we can take the trace over all such ingoing modes. This gives the reduced density operator for the right moving modes as
\begin{eqnarray}
{\hat{\rho}}^{(R)}_{(\textrm{boson})}= \Big(1-e^{-\frac{2\pi\Omega}{\hbar\kappa}}\Big) \displaystyle\sum_{n} e^{-\frac{2\pi n\Omega}{\hbar\kappa}}|n^{(R)}_{\textrm{obs}}\rangle \langle n^{(R)}_{\textrm{obs}}|
\label{densityright}
\end{eqnarray}
The average number of particles detected at asymptotic infinity, given by the expectation value of the number operator $\hat{n}$, is now given by
\begin{eqnarray}
\langle n\rangle_{(\textrm{boson})} &=& {\textrm{trace}}({\hat{n}} {\hat{\rho}}^{(R)}_{(\textrm{boson})})\nonumber\\
                       &=& \frac{1}{e^{\frac{2\pi\Omega}{\hbar\kappa}}-1},
\label{expectation}
\end{eqnarray}
which is nothing but the Bose-Einstein distribution of particles corresponding to the Hawking temperature 
\begin{eqnarray}
T_{\textrm H}=\frac{\hbar\kappa}{2\pi}=\frac{\hbar g'(r_h)}{4\pi}.
\label{semihawk}
\end{eqnarray}
The same methodology, when applied on fermions, gives the Fermi-Dirac distribution with the correct Hawking temperature. In the context of Lovelock black holes with distinct asymptotic behavior, we now substitute $g(r)$ by the appropriate metric coefficients to reproduce the known semiclassical Hawking temperature for each spacetime.

Now it may be worthwhile to mention that one could have chosen the set of transformations (\ref{connect1}) with opposite relative sign between two terms at the right hand side, so that, 
\begin{eqnarray}
t_{in} &=& t_{obs}+\frac{i\pi}{2\kappa}\nonumber\\
r^*_{in} &=&  r^*_{obs}-\frac{i\pi}{2\kappa}.
\label{alter}
\end{eqnarray}
For this choice also the inside and outside coordinates (\ref{transformation}), (\ref{transformation2}) are connected. However this is an unphysical solution \cite{Majhivag}. To see this note that, use of (\ref{alter}), gives 
\begin{eqnarray}
\Phi^L_{in} &=& \Phi^L_{obs}\nonumber\\
\Phi^R_{in} &=& e^\frac{\pi\Omega}{\hbar\kappa}\Phi^R_{obs}.
\label{alter1}
\end{eqnarray} 
Therefore the probabilities, that the ingoing (left-moving) modes can go inside the event horizon ($P^{R}$) and the outgoing (right-moving) modes can go outside the event horizon, as observed from outside,  are given by
\begin{eqnarray}
P^L=|\Phi^L_{in}|^2=|\Phi^L_{obs}|=1\nonumber\\
P^R=|\Phi^R_{in}|^2=e^\frac{2\pi\Omega}{\hbar\kappa}|\Phi^R_{obs}|^2=e^\frac{2\pi\Omega}{\hbar\kappa}.
\label{alter2}
\end{eqnarray} 
In the classical limit ($\hbar\rightarrow 0$), there is absolutely no chance that any mode can cross the event horizon from inside, therefore one must have $P^R=0$. But we can see from (\ref{alter2}) that it is diverging and therefore the choice (\ref{alter}) is unphysical. It may be worthwhile to mention that, interestingly, for the cosmological horizon a choice similar to (\ref{alter}) will be physical whereas the other choice similar to (\ref{connect1}) is unphysical.

\subsection{Blackbody spectrum for the cosmological horizon}
To find the radiation spectrum for the cosmological horizon we first need to perform two tasks. One is the Kruskal-like extension of the space-time (\ref{metric}) just around the cosmological horizon which is done in appendix (\ref{sec:B}). The other requirement is to identify the left and right moving modes outside and inside the cosmological horizon which is discussed in appendix (\ref{sec:C}) {\footnote{We use the subscript ``$obs$'' for modes inside the cosmological horizon, such that $r_h<r<r_c$, since an observer can stay only in this region. The subscript ``$out$'' is used for modes outside the cosmological horizon.}}. 

From the Kruskal-like extension it is found that the inside ($T_{obs}, X_{obs}$) and outside ($T_{out}, X_{out}$) coordinates, defined in (\ref{transformation4}) and (\ref{transformation3}) respectively, can be connected with each other by the following relations {\footnote{One can again choose the opposite relative signs between the quantities at the right hand side of (\ref{cosconnect1}). With this choice the probability for a right-moving mode to cross the cosmological horizon from inside is $P^R=1$. The probability for the left-moving mode to cross the cosmological horizon from the outside, observed from inside the horizon, is then given by $P^L=e^\frac{-2\pi\Omega}{\hbar\kappa}$. Since, for the cosmological horizon, $\Omega$ is the BBM energy which is negative \cite{BBM}, $P^L$ diverges in the classical limit ($\hbar\rightarrow 0$). Therefore (\ref{cosconnect1}) is the only physical choice for this case.}}
{\begin{eqnarray}
t_{out} &=& t_{obs}+\frac{i\pi}{2\kappa}\nonumber\\
r^*_{out} &=&  r^*_{obs}-\frac{i\pi}{2\kappa}.
\label{cosconnect1}
\end{eqnarray} 
The left and right moving modes inside the cosmological horizon (outside the event horizon) are given by (\ref{conv5}) and (\ref{conv6}) respectively, whereas, (\ref{conv7}) and (\ref{conv8}) give left and right moving modes outside the cosmological horizon respectively. Using (\ref{cosconnect1}) the modes inside and outside the cosmological horizon can be connected as
\begin{eqnarray}
\Phi_{out}^{(R)} &=& \Phi_{{obs}}^{{(R)}} \nonumber\\
\Phi_{out}^{(L)} &=& {e^{\frac{\pi\Omega}{\hbar\kappa}}}{\Phi_{{obs}}^{{(L)}}}
\label{cosconnect3}
\end{eqnarray}
The physical state representing $n$ number of non-interacting pairs, created outside the cosmological horizon, when viewed from inside the horizon, is given by
\begin{eqnarray}
|\Psi\rangle = N \displaystyle\sum_n |n^{(R)}_{\textrm{out}}\rangle\otimes|n^{(L)}_{\textrm{out}}\rangle
 = N \displaystyle\sum_n e^{\frac{\pi n\Omega}{\hbar\kappa}}|n^{(R)}_{\textrm{obs}}\rangle\otimes|n^{(L)}_{\textrm{obs}}\rangle.
\label{cosstate}
\end{eqnarray} 
Here the normalization constant $N$ can be found from $\langle\Psi|\Psi\rangle=1$ and the physical state for bosons and fermions, turns out to be,
\begin{eqnarray}
|\Psi\rangle_{(\textrm{boson})}= \Big(1-e^{\frac{2\pi\Omega}{\hbar\kappa}}\Big)^{\frac{1}{2}} \displaystyle\sum_n e^{\frac{\pi n\Omega}{\hbar\kappa}}|n^{(R)}_{\textrm{obs}}\rangle\otimes|n^{(L)}_{\textrm{obs}}\rangle,
\label{cosboson}
\\
|\Psi\rangle_{(\textrm{fermion})}= \Big(1+e^{\frac{2\pi\Omega}{\hbar\kappa}}\Big)^{-\frac{1}{2}} \displaystyle\sum_n e^{\frac{\pi n\Omega}{\hbar\kappa}}|n^{(R)}_{\textrm{obs}}\rangle\otimes|n^{(L)}_{\textrm{obs}}\rangle
\label{cosfermion}
\end{eqnarray}   
respectively. The density operator for the bosons is now constructed as 
\begin{eqnarray}
{\hat\rho}_{(\textrm{boson})}&=&|\Psi\rangle_{(\textrm{boson})}\langle\Psi|_{(\textrm{boson})}
\nonumber
\\
&=&\Big(1-e^{\frac{2\pi\Omega}{\hbar\kappa}}\Big) \displaystyle\sum_{n,m} e^{\frac{\pi (n+m)\Omega}{\hbar\kappa}} |n^{(R)}_{\textrm{obs}}\rangle\otimes|n^{(L)}_{\textrm{obs}}\rangle\langle m^{(R)}_{\textrm{obs}}|\otimes\langle m^{(L)}_{\textrm{obs}}|.
\label{cosdens}
\end{eqnarray}
Since in this case right-moving modes are going outside the cosmological horizon, these are completely lost. We take the the trace over all such right-moving modes to find the reduced density operator for the left-moving modes, given by 
\begin{eqnarray}
{\hat{\rho}}^{(L)}_{(\textrm{boson})}= \Big(1-e^{\frac{2\pi\Omega}{\hbar\kappa}}\Big) \displaystyle\sum_{n} e^{\frac{2\pi n\Omega}{\hbar\kappa}}|n^{(L)}_{\textrm{obs}}\rangle \langle n^{(L)}_{\textrm{obs}}|.
\label{cosdenright}
\end{eqnarray}
In the case of cosmological horizon the particles are not observed at asymptotic infinity, rather in a region in between the event and the cosmological horizon. The average number of particles which is detected by an observer in this region is now given by, 
\begin{eqnarray}
\langle n\rangle_{(\textrm{boson})} &=& {\textrm{trace}}({\hat{n}} {\hat{\rho}}^{(L)}_{(\textrm{boson})})\nonumber\\
                       &=& \frac{1}{e^{-\frac{2\pi\Omega}{\hbar\kappa}}-1}.
\label{cosexp}
\end{eqnarray}
This is again a Bose-Einstein distribution of particles corresponding to the new Hawking temperature 
\begin{eqnarray}
T_{\textrm c}=-\frac{\hbar\kappa}{2\pi}=-\frac{\hbar g'(r_c)}{4\pi}.
\label{semicos}
\end{eqnarray}
Note that the temperature has the same value (\ref{semihawk}) as found for the black hole (event) horizon but is negative. This negative temperature together with the negative BBM energy \cite{BBM} of the dS spacetime make the first law of thermodynamics valid for the cosmological horizon.


\section{Blackbody spectrum and corrected Hawking temperature}{\label{sec:4}}

We now develop a general framework to find the corrections to the semiclassical Hawking radiation for a general class of metric (\ref{metric}),  where the $r-t$ sector of the metric is decoupled from the angular parts. In our previous work \cite{Modak} we generalized the tunneling method to find the corrections to the Hawking temperature for the black hole solution of Einstein-Maxwell theory (Kerr-Newman) in (3+1) dimensions by using the method of complex path. Now we shall use the modified method \cite{Majhiflux}, as outlined in the previous section, where one can find the radiation spectrum. We now generalize the approach presented in section 3, by going beyond the semiclassical approximation, to find the modified radiation spectrum. Our analysis shows that the higher order terms in the WKB ansatz, when included in the theory, do not affect the thermal nature of the spectrum. Grey-body factors do not appear in the radiation spectrum, rather the temperature of the radiation undergoes some quantum corrections, with a perfect blackbody spectrum.

In the beginning of section 3 we only considered the semiclassical action (${\cal S}_{0}$) corresponding to the scalar field ansatz (\ref{2.2}) and found a solution for that in (\ref{soln}). Now we want to follow the approach developed in \cite{Majhibeyond} to include all the higher order terms in $\hbar$ in the analysis. Putting (\ref{2.2}) in (\ref{2.1}) and equating the coefficients of different orders in $\hbar$ to zero one finds a set of partial differential equations \cite{Majhibeyond}. Each differential equation corresponding to any specific power of $\hbar$ can be simplified by the equation coming in one lower order in $\hbar$. This finally yields \cite{Modak,Majhibeyond,Modak1,Majhitrace,group,Tao} the set of partial differential equations for different powers of $\hbar$,
\begin{eqnarray}
\hbar^0~:~\frac{\partial S_0}{\partial t}=\pm g(r)\frac{\partial S_0}{\partial r},
\label{2.5a}
\end{eqnarray}
\begin{eqnarray}
\hbar^1~:~&&\frac{\partial S_1}{\partial t}=\pm g(r)\frac{\partial S_1}{\partial r},
\nonumber
\\
\hbar^2~:~&&\frac{\partial S_2}{\partial t}=\pm g(r)\frac{\partial S_2}{\partial r},
\nonumber
\\
.
\nonumber
\\
.
\nonumber
\\
.
\nonumber
\end{eqnarray} 
and so on. Therefore the $n$-th order solution of (\ref{2.3}) is given by,  
\begin{eqnarray}
\frac{\partial S_n}{\partial t}=\pm g(r)\frac{\partial S_n}{\partial r},
\label{2.5}
\end{eqnarray}
where ($n=~0,~i;~i= 1,2,...)$.

Because of this identical set of differential equations, the solutions for other ${\cal S}_i(r,t)$'s, subjected to a similar functional choice like (\ref{actionform}), can differ only by a proportionality factor from (\ref{action2}). Therefore the most general form of ${\cal S}(r,t)$ can be written as 
\begin{eqnarray}
{\cal S}(r,t)=(1+\displaystyle\sum_{i=1}^{\infty}\gamma_i \hbar^i) {\cal S}_0(r,t),
\label{2.7}
\end{eqnarray}
where $\gamma_i$'s are proportionality constants having dimensions of $[\hbar]^{-i}$.

We carry out the following dimensional analysis to express these $\gamma_i$'s in terms of dimensionless constants. The fact that the units of the product of Newton's constant and mass density are same in all dimensions \cite{Zwiebach} yields
\begin{eqnarray}
[G]\frac{M}{L^3}=[G^{(D)}]\frac{M}{L^{D-1}},
\label{2.8}
\end{eqnarray}
where $G$ is the Newton's constant in four dimensions, given by 
\begin{eqnarray}
[G]=\frac{[c]^{3}L^{2}}{[\hbar]}.
\label{2.9}
\end{eqnarray}
Putting this in (\ref{2.8}) and replacing $L$ by $D$-dimensional Planck length ($l_P^{(D)}$) we get  
\begin{eqnarray}
\left(l_P^{(D)}\right)^{(D-2)}=\frac{[\hbar]G^{(D)}}{[c]^3}.
\label{2.10}
\end{eqnarray}
In units of $G^{(D)}=c=1$, one finds $[\hbar]=\left(l_P^{(D)}\right)^{(D-2)}$. The only black hole parameter having the unit of length is the horizon radius ($r_h$ or $r_c$). Therefore we can write $[\hbar]\sim r_h^{(D-2)}$ or $[\hbar]\sim r_c^{(D-2)}$ for the black hole (event) horizon and the cosmological horizon respectively. Now (\ref{2.7}) can be written as (for the tunneling through the black hole horizon)
\begin{eqnarray}
{\cal S}(r,t) &=& \left(1+\displaystyle\sum_{i=1}^{\infty}\frac{\beta_i \hbar^i}{r_h^{i(D-2)}}\right){\cal S}_0(r,t)\nonumber\\
              &=& \left(1+\displaystyle\sum_{i=1}^{\infty}\frac{\beta_i \hbar^i}{r_h^{i(D-2)}}\right) \left(\Omega t \pm \Omega\int\frac{dr}{g(r)}\right),
\label{2.11}
\end{eqnarray}
where $\beta_i$'s are dimensionless constants. Finally the cherished solution for the scalar field in presence of the higher order corrections to the semiclassical action, follows from (\ref{2.2}) and (\ref{2.11}),
\begin{eqnarray}
\Phi=\exp[-{\frac{i}{\hbar}}\Omega\left(1+\displaystyle\sum_{i=1}^{\infty}\frac{\beta_i \hbar^i}{r_h^{i(D-2)}}\right)(t\pm r^{*})].
\label{highsoln}
\end{eqnarray}

The left and right moving modes inside and outside the black hole event horizon, following the convention of appendix (\ref{sec:C}), now becomes
\begin{eqnarray}
&&\Phi_{in}^{(R)} = e^{-\frac{i}{\hbar}\Omega\left(1+\displaystyle\sum_{i=1}^{\infty}\frac{\beta_i \hbar^i}{r_h^{i(D-2)}}\right) u_{in}};\,\,\ \Phi_{in}^{(L)}=e^{-\frac{i}{\hbar}\Omega\left(1+\displaystyle\sum_{i=1}^{\infty}\frac{\beta_i \hbar^i}{r_h^{i(D-2)}}\right) v_{in}},
\nonumber
\\
&&\Phi_{out}^{(R)}=e^{-\frac{i}{\hbar}\Omega\left(1+\displaystyle\sum_{i=1}^{\infty}\frac{\beta_i \hbar^i}{r_h^{i(D-2)}}\right) u_{out}};\,\,\,\Phi_{out}^{(L)}=e^{-\frac{i}{\hbar}\Omega\left(1+\displaystyle\sum_{i=1}^{\infty}\frac{\beta_i \hbar^i}{r_h^{i(D-2)}}\right) v_{out}}.
\label{highmodes}
\end{eqnarray}
These inside and outside modes, moving in a particular direction (right or left), can also be connected by the set of transformations (\ref{connect1}) or (\ref{connect2}). This yields
\begin{eqnarray}
\Phi_{in}^{(R)} &=& {e^{-\frac{\pi\Omega}{\hbar\kappa}\left(1+\displaystyle\displaystyle\sum_{i=1}^{\infty}\frac{\beta_i \hbar^i}{r_h^{i(D-2)}}\right)}}\Phi_{out}^{(R)}\nonumber\\
 &=& {e^{-\frac{\pi\Omega}{\hbar{\kappa}'}}}\Phi_{out}^{(R)}\nonumber\\
\Phi_{in}^{(L)} &=& \Phi_{out}^{(L)},
\label{highconnect}
\end{eqnarray}
where we have substituted 
\begin{eqnarray}
\kappa '=\left(1+\displaystyle\displaystyle\sum_{i=1}^{\infty}\frac{\beta_i \hbar^i}{r_h^{i(D-2)}}\right)^{-1}\kappa.
\label{highsurface}
\end{eqnarray}
This can be considered as the modified surface gravity in presence of higher $\hbar$ order corrections to the WKB ansatz. 

The physical state representing $n$ number of virtual pairs inside the horizon, when observed from the outside, is now given by
\begin{eqnarray}
|\Psi\rangle = N \displaystyle\sum_n |n^{(L)}_{\textrm{in}}\rangle\otimes|n^{(R)}_{\textrm{in}}\rangle
 = N \displaystyle\sum_n e^{-\frac{\pi n\Omega}{\hbar\kappa '}}|n^{(L)}_{\textrm{out}}\rangle\otimes|n^{(R)}_{\textrm{out}}\rangle.
\label{highstate}
\end{eqnarray}
Now from here on, it is trivial to check that the whole methodology developed for the semiclassical case can be repeated to find the new radiation spectrum. The only difference is the redefinition of the surface gravity ($\kappa$) by $\kappa '$. The final result for the radiation spectrum, for bosons, is now given by
\begin{eqnarray}
\langle n\rangle_{(\textrm{boson})}= \frac{1}{e^{\frac{2\pi\Omega}{\hbar\kappa '}}-1}.
\label{highexpect}
\end{eqnarray}     
Now one can see that the spectrum is still given by the blackbody spectrum with the new corrected Hawking temperature  
\begin{eqnarray}
T_{\textrm{bh}} &=& \frac{\hbar\kappa '}{2\pi}=\frac{\hbar\kappa}{2\pi}\left(1+\displaystyle\displaystyle\sum_{i=1}^{\infty}\frac{\beta_i \hbar^i}{r_h^{i(D-2)}}\right)^{-1}\nonumber\\
                &=& \left(1+\displaystyle\displaystyle\sum_{i=1}^{\infty}\frac{\beta_i \hbar^i}{r_h^{i(D-2)}}\right)^{-1}T_{\textrm{H}},
\label{correcttemp}
\end{eqnarray}
where $T_{\textrm{H}}$ is the usual semiclassical Hawking temperature, given by (\ref{semihawk}). Note that (\ref{correcttemp}) gives the corrected Hawking temperature for any general static, chargeless black hole solutions in EGB theory with an appropriate choice of metric.

For the cosmological horizon, in presence of other higher order terms in $\hbar$ in the action, the relation between right and left moving modes at two sides are given by 
\begin{eqnarray}
\Phi_{out}^{(R)} &=& \Phi_{{obs}}^{{(R)}} \nonumber\\
\Phi_{out}^{(L)} &=& {e^{\frac{\pi\Omega}{\hbar\kappa '}}}{\Phi_{{obs}}^{{(L)}}},
\label{cosconnect3}
\end{eqnarray}
where $\kappa '$ is defined in (\ref{highsurface}). Subsequently the new radiation spectrum for the bosons turns out to be same as (\ref{expectation}) with $\kappa '$ replacing $\kappa$. Therefore the modified Hawking temperature for the cosmological horizon, given by replacing $r_h$ by $r_c$, is
\begin{eqnarray}
T_{\textrm{ch}} &=& -\frac{\hbar\kappa '}{2\pi}=\frac{\hbar\kappa}{2\pi}\left(1+\displaystyle\displaystyle\sum_{i=1}^{\infty}\frac{\beta_i \hbar^i}{r_c^{i(D-2)}}\right)^{-1}\nonumber\\
                &=& \left(1+\displaystyle\displaystyle\sum_{i=1}^{\infty}\frac{\beta_i \hbar^i}{r_c^{i(D-2)}}\right)^{-1}T_{\textrm{c}},
\label{corcosttemp}
\end{eqnarray}
where $T_{\textrm c}$ is the semiclassical temperature given by (\ref{semicos}).

The coefficients $\beta_i$ s' occurring in either (\ref{correcttemp}) or (\ref{corcosttemp}) are related to the trace anomaly. This will be discussed in section 6.


\subsection{Spherically symmetric, asymptotically flat black holes}

For these black holes $g(r_h)=f(r_h)$, where $f(r)$ is given by (\ref{1.6}). Therefore from (\ref{correcttemp}), the corrected Hawking temperature for the spherically symmetric, asymptotically flat Lovelock black holes is given by
\begin{eqnarray}
T_{\textrm {bh}}=\frac{\hbar (D-3)}{4\pi r_h}\Big(1+\displaystyle\sum_i\beta_i\frac{\hbar^i}{r_h^{i(D-2)}}\Big)^{-1}\left[\frac{r_h^2+\frac{D-5}{D-3}\frac{\bar\lambda}{2}}{r_h^2+\bar\lambda}\right].
\label{2.17a}
\end{eqnarray}
The semiclassical Hawking temperature is given by the leading contribution,
\begin{eqnarray}
T_{\textrm H}=\frac{\hbar f'(r_h)}{4\pi}=\frac{\hbar (D-3)}{4\pi r_h}\left[\frac{r_h^2+\frac{D-5}{D-3}\frac{\bar\lambda}{2}}{r_h^2+\bar\lambda}\right],
\label{2.17}
\end{eqnarray}
This is in agreement with (\ref{1.8}), present in the existing literature \cite{Myers}, found by following Hawking's original derivation to avoid the conical singularity at the black hole event horizon.


\subsection{Topological Gauss-Bonnet AdS black holes}

One can calculate the modified Hawking temperature for the AdS black hole of section 2.2 just by replacing $g(r_h)=F(r_h)$ from (\ref{1.11}). This would lead to the result, with $r_h$ now representing the event horizon radius for AdS black holes, for the black hole temperature, 
\begin{eqnarray}
T_{\textrm{bh}}^{\textrm{(AdS)}}=\Big(1+\displaystyle\sum_i\beta_i\frac{\hbar^i}{r_h^{i(D-2)}}\Big)^{-1}\frac{\hbar[(D-1)r_h^4+D-3)kl^2r^2+(D-5)\tilde\alpha k^2l^2]}{4\pi l^2r_h(r_h^2+2\tilde\alpha k)}.
\label{2.19}
\end{eqnarray}
Here also at the lowest order in $\hbar$, we get the known value of the semiclassical Hawking temperature (\ref{1.12}).  


\subsection{Gauss-Bonnet dS black holes}

For the dS black holes in section 2.3 the issue of Hawking radiation is quite subtle because there is no notion of spatial infinity as spacetime is bounded by a cosmological horizon. In a work \cite{Teitel} Gomberoff and Teitelboim argued that in the absence of spatial infinity one can take either the black hole horizon or the cosmological horizon as the boundary to study the thermodynamics of the other horizon. The black hole horizon is always inside the cosmological horizon. It is known that both the horizons are involved in Hawking radiation and the observer is somewhere in between the two horizons. The Hawking temperature associated with these two horizons are not equal. As a result there is no equilibrium between them, if they are  treated as two different systems. Ideally any observer in between the two horizons will get the radiation coming from both the horizons. Unlike the black hole event horizon, the pair creation occurs just outside of the cosmic horizon. The outgoing mode goes away from the cosmic horizon and the other one tunnels inwards to reach the observer.

The corrected Hawking temperature associated with the {\it black hole horizon}, found by using $g(r_h)={\tilde F(r_h)}$, is given by
\begin{eqnarray}
T_{\textrm {bh}}^{\textrm{(dS)}}=T_{\textrm H}^{\textrm{(dS)}}\Big(1+\displaystyle\sum_i\beta_i\frac{\hbar^i}{r_h^{i(D-2)}}\Big)^{-1},
\label{2.20}
\end{eqnarray}  
where the semiclassical Hawking temperature ($T_{\textrm H}^{\textrm{(dS)}}$) agrees with (\ref{1.18}).

For the cosmological horizon it is given by (\ref{corcosttemp}) and substituting $g(r_c)=\tilde F(r_c)$ we find
\begin{eqnarray}
T_{\textrm {ch}}^{\textrm{(dS)}}=T_{\textrm c}\Big(1+\displaystyle\sum_i\beta_i\frac{\hbar^i}{r_c^{i(D-2)}}\Big)^{-1},
\label{2.21}
\end{eqnarray}    
where the semiclassical value $T_c$ exactly matches with $T_c^{\textrm{dS}}$ in (\ref{1.20}).

\section{Entropy Correction}{\label{sec:5}}

The fact that the higher order corrections to the WKB ansatz yields quantum corrections to the black hole temperature enable us to find the entropy corresponding to that modified temperature. While doing so we shall use the basic assumption that the Lovelock black holes satisfy the first law of thermodynamics $dM=TdS$. Then one can find the  corrected entropy of Lovelock black holes corresponding to the corresponding corrected Hawking temperature as
\begin{eqnarray}
S_{\textrm{bh}}=\int\frac{dM}{T_{\textrm{bh}}}=\int{\frac{1}{T_{\textrm{bh}}}\left(\frac{\partial M}{\partial r_h}\right)dr_h}.
\label{3.1}
\end{eqnarray}
The same method will work for the cosmological horizon (for the Gauss-Bonnet dS black holes), where we shall use the modified Hawking temperature of cosmological horizon. It may be mentioned that, following this thermodynamical approach, the semiclassical entropy was computed in \cite{Cai} and \cite{Cai2}. Although it is possible to find the correctional terms for all orders, here we shall concentrate only up-to leading and sub-leading contributions.

\subsection{Spherically symmetric, asymptotically flat black holes}

The gravitational mass and corrected Hawking temperature, expressed in terms of the horizon radius, are given by (\ref{1.7a}) and (\ref{2.17a}) respectively. Using these in (\ref{3.1}) and integrating up-to second order in $\hbar$, we get

\begin{eqnarray}
S_{\textrm{bh}}^{\textrm{flat}}=\frac{A_{D-2}r_h^{(D-2)}}{4G\hbar}\left(1+\frac{D-2}{D-4}\frac{\bar\lambda}{r_h^2}\right)+\frac{\beta_1A_{D-2}}{4G}\left(\log(\frac{r_h^{D-2}}{G\hbar}) -\frac{\bar\lambda(D-2)}{2r_h^2}\right)\nonumber\\
-\frac{\beta_2\hbar A_{D-2}}{4G}\left(\frac{1}{r_h^{(D-2)}}+(1-\frac{2}{D})\frac{\bar\lambda}{r_h^D}\right)+{\cal O}(\hbar^2). 
\label{3.2}
\end{eqnarray}
The first term in this expression is the familiar semiclassical entropy for the spherically symmetric black hole solutions in Lovelock gravity \cite{Myers} and other terms are coming as quantum corrections. The semiclassical result for black hole entropy is not just one-quarter of its horizon area. Because of the presence of the Gauss-Bonnet term it is modified upto an additive term which also can be written as a function of horizon area. From (\ref{3.2}) it is clear that the leading order correction to the semiclassical value is not pure logarithmic and also the next to leading correction is not inverse of the horizon area. This is in contrast to Einstein's gravity where for the Schwarzschild solution, one has purely logarithmic and inverse horizon area terms as the leading and next to leading order corrections, respectively. The presence of Gauss-Bonnet term, therefore, modifies the results in a nontrivial manner.

From (\ref{3.2}) it follows that in the limit $\lambda=0$ (i.e. without the Gauss-Bonnet term) we get
\begin{eqnarray}
S_{\textrm{bh}}(\lambda=0)=S_{\textrm{BH}}+\frac{\beta_1(D-2)A_{D-2}}{4G}\log{S_{\textrm{BH}}}-\beta_2{(\frac{A_{D-2}}{4G})}^2\frac{1}{S_{\textrm{BH}}}\nonumber\\
+{\cal O}(\hbar^2). 
\label{3.3}
\end{eqnarray}
This is the corrected entropy for the Schwarzschild solution in Einstein gravity in arbitrary $D$- dimensions and here one has logarithmic and inverse area terms as leading and subleading corrections. In particular, for $D=4$, (\ref{3.3}) reproduces the result of corrected entropy for the Schwarzschild black hole, found in tunneling \cite{ Modak, Majhibeyond} or path integral \cite{Hawking2} approaches, where the coefficient of the logarithmic term is given by $\frac{1}{90}$ (in $c=G=\kappa_B=1$ unit).

\subsection{Topological Gauss-Bonnet AdS black holes}
The corrected entropy for these black holes can be calculated by integrating the first law of thermodynamics (\ref{3.1}) with the modified temperature (\ref{2.19}) and black hole mass (\ref{1.11a}), which yields
\begin{eqnarray}
S_{\textrm{bh}}^{\textrm{AdS}}=\frac{\Sigma_kr_h^{(D-2)}}{4G\hbar}\left(1+\frac{D-2}{D-4}\frac{2\tilde\alpha k}{r_h^2}\right)+\frac{\beta_1\Sigma_{k}}{4G}\left(\log(\frac{r_h^{D-2}}{G\hbar}) -\frac{(D-2)\tilde\alpha k}{r_h^2}\right)\nonumber\\
-\frac{\beta_2\hbar \Sigma_{k}}{4G}\left(\frac{1}{r_h^{(D-2)}}+(1-\frac{2}{D})\frac{2\tilde\alpha k}{r_h^D}\right)+{\cal O}(\hbar^2). 
\label{3.4}
\end{eqnarray}

Now it is interesting to see that for $k=0$ the functional form of corrected entropy (\ref{3.4}) is identical to the result found for Einstein gravity (\ref{3.3}), although the two cases differ drastically. For $k=0$ one has a zero curvature hypersurface representing an event horizon and for the second case one does not have the Gauss-Bonnet term in the theory. Previously, this particular phenomena was known for the semiclassical case only, but here we can see the same is true also for the corrected form of entropy.

\subsection{Gauss-Bonnet dS black holes}

For the dS black holes we have to find the correctional terms to the semiclassical entropy separately for the black hole (event) horizon and the cosmological horizon.

\subsubsection{Black hole horizon}

Using the gravitational mass or the AD mass of the black hole from (\ref{1.17a}), the modified Hawking temperature from (\ref{2.20}) and integrating the first law of thermodynamics (\ref{3.1}) we get
\begin{eqnarray}
S_{\textrm{bh}}^{\textrm{dS}}=\frac{A_{D-2}r_h^{(D-2)}}{4G\hbar}\left(1+\frac{D-2}{D-4}\frac{2\tilde\alpha }{r_h^2}\right)+\frac{\beta_1A_{D-2}}{4G}\left(\log(\frac{r_h^{D-2}}{G\hbar}) -\frac{(D-2)\tilde\alpha}{r_h^2}\right)\nonumber\\
-\frac{\beta_2\hbar A_{D-2}}{4G}\left(\frac{1}{r_h^{(D-2)}}+(1-\frac{2}{D})\frac{2\tilde\alpha}{r_h^D}\right)+{\cal{O}}(\hbar^2). 
\label{3.5}
\end{eqnarray}

\subsubsection{Cosmological horizon}
The gravitational mass corresponding to the cosmological horizon, given by the BBM prescription \cite{BBM}, is just negative of the AD mass (\ref{1.17a}) with $r_c$ replacing $r_h$. Now using (\ref{2.21}) in (\ref{3.1}) (with $T_{\textrm{ch}}$ replacing $T_{\textrm{bh}}$), one can calculate the corrected entropy associated with the cosmological horizon,
\begin{eqnarray}
S_{\textrm{ch}}^{\textrm{dS}}=\frac{A_{D-2}r_c^{(D-2)}}{4G\hbar}\left(1+\frac{D-2}{D-4}\frac{2\tilde\alpha }{r_c^2}\right)+\frac{\beta_1A_{D-2}}{4G}\left(\log(\frac{r_c^{D-2}}{G\hbar}) -\frac{(D-2)\tilde\alpha }{r_c^2}\right)\nonumber\\
-\frac{\beta_2\hbar A_{D-2}}{4G}\left(\frac{1}{r_c^{(D-2)}}+(1-\frac{2}{D})\frac{2\tilde\alpha}{r_c^D}\right)+{\cal O}(\hbar^2). 
\label{3.6}
\end{eqnarray}


\section{Trace anomaly and coefficient of the leading correction}{\label{sec:6}}

In this section we relate the coefficient ($\beta_1$) of the leading correction to entropy for various cases (\ref{3.2}), (\ref{3.4}), (\ref{3.5}) and (\ref{3.6}) with the trace anomaly of the stress tensor for the scalar field, moving in a background of `$D$' dimensional curved spacetime. For that we use the method of complex path proposed in \cite{Paddy} and which has been used earlier for the same purpose in \cite{Modak,Majhitrace}. In this method particle creation occurs just inside the black hole horizon (just outside the cosmological horizon(CH)). One mode is attracted towards the center of a black hole and the outgoing mode tunnels through the black hole (event) horizon, traversing a complex path which is forbidden classically (for CH the outgoing mode goes away while the ingoing mode tunnels inside). In our convention the $+$ ($-$) sign in (\ref{2.11}) or (\ref{highsoln}) implies that the particle is ingoing (outgoing). The expression for the ingoing and outgoing scalar field modes, following from (\ref{highsoln}), are then given by
\begin{eqnarray}
\Phi_{in} &=& \exp[-{\frac{i}{\hbar}}\Omega\left(1+\displaystyle\sum_i\frac{\beta_i \hbar^i}{r_h^{i(D-2)}}\right)(t+\int_C\frac{dr}{g(r)})]\nonumber\\
\Phi_{out} &=& \exp[-{\frac{i}{\hbar}}\Omega\left(1+\displaystyle\sum_i\frac{\beta_i \hbar^i}{r_h^{i(D-2)}}\right)(t-\int_C\frac{dr}{g(r)})]\label{trace1}
\end{eqnarray}
For the black hole event horizon the contour is chosen such that it starts just behind the event horizon left to right along a semicircle, in the lower half of the complex plane, just avoiding the singularity at the (event) horizon. The ingoing and outgoing probabilities for the respective modes in the case of black hole (event) horizon are given by
\begin{eqnarray}
P_{in} &=& {\exp}\Big[\frac{2}{\hbar}(1+\sum_i\beta_i\frac{\hbar^i}{r_h^{i(D-2)}})\Big(\Omega{\textrm{Im}}~t +\Omega{\textrm{Im}}\int_C\frac{dr}{g(r)}\Big)\Big]\nonumber\\
P_{out} &=& {\exp}\Big[\frac{2}{\hbar}(1+\sum_i\beta_i\frac{\hbar^i}{r_h^{i(D-2)}})\Big(\Omega{\textrm{Im}}~t -\Omega{\textrm{Im}}\int_C\frac{dr}{g(r)}\Big)\Big].\label{trace2}
\end{eqnarray}
Since classically the ingoing probability $P_{in}=1$, one has 
\begin{eqnarray}
{\textrm{Im}~t}=-{\textrm{Im}}\int_C\frac{dr}{g(r)}
\label{trace3}
\end{eqnarray}
for the black hole horizon. 

Now let us make an infinitesimal scale transformation to the metric coefficient of the Lovelock spcetime as $\bar{g_{tt}}(r)=kg_{tt}(r)\simeq(1+\delta k)g_{tt}(r)$ and $\bar g^{rr}(r)=k^{-1}g^{rr}(r)\simeq(1+\delta k)^{-1}g^{rr}(r)$. The invariance of the Klein-Gordon equation under this transformation enforces the field ($\Phi$) to transform as $\bar\Phi=k^{-1}\Phi$. Now consider the action for the massive scalar field in arbitrary `$D$' dimensions, given by
\begin{eqnarray}
{\cal {S}}=\frac{1}{2}\int{\sqrt{-g}(\nabla_{\mu}\Phi\nabla^{\mu}\Phi-m^2\Phi^2)d^{D}x},
\label{action3}
\end{eqnarray}  
where $m$ is the mass of the scalar field{\footnote{We keep $m$ here only for the dimensional analysis, otherwise it is always zero in our analysis which considers the tunneling of massless scalar fields.}}. From this one finds that the dimension of $\Phi$ is given by, 
\begin{eqnarray}
[\Phi]\sim[M]^{(D-2)/2}\sim\frac{1}{[L]^{(D-2)/2}}. 
\label{dimension}
\end{eqnarray}
Comparing the scalar particle action from (\ref{action}) and (\ref{2.11}) one has
\begin{eqnarray}
{\textrm{Im}}{\cal{S}}^{\textrm{out}}_{1}(r,t)=\frac{\beta_1}{r_h^{(D-2)}}{\textrm{Im}}{\cal{S}}^{\textrm{out}}_{0}(r,t). 
\label{compare}
\end{eqnarray}
Now concentrating on the tunneling from the black hole (event) horizon and using (\ref{actionform}), (\ref{s0}) and (\ref{trace3}), we get
\begin{eqnarray}
{\textrm{Im}}{\cal{S}}^{\textrm{out}}_{0}(r,t)= -2\Omega{\textrm{Im}}\int_C{\frac{dr}{g(r)}}\label{ims0}
\end{eqnarray}


Under scale transformation the scalar field transforms as $\bar\Phi=k^{-1}\Phi$. Therefore from (\ref{dimension}) it follows that the mass and length scales  should be changed as $\bar M= k^{-2/(D-2)}M\simeq(1-\frac{2}{D-2}\delta k)M$ and ${(1/\bar L)}= k^{-2/(D-2)}{(1/L)}\simeq(1-\frac{2}{D-2}\delta k){(1/L)}$ respectively. The same is true for the gravitational energy ($\Omega$) and horizon radius ($r_h$ or $r_c$) of the Lovelock spacetime in arbitrary `$D$' dimensions. From these we can calculate the transformed form of (\ref{compare}). This is found to be,
\begin{eqnarray}
{\textrm{Im}}\bar{\cal{S}}^{\textrm{out}}_{1}(r,t)=\frac{\beta_1}{\bar{\textrm{r}}^{(D-2)}_{h}}{\textrm{Im}}\bar{\cal{S}}^{\textrm{out}}_{0}(r,t)\simeq\frac{\beta_1}{{r_\textrm{h}}^{(D-2)}}(1-\frac{4}{D-2}\delta k){\textrm{Im}{\cal{S}}^{\textrm{out}}_{0}(r,t)}.
\label{transformed}
\end{eqnarray}
which leads to  
\begin{eqnarray}
\frac{\delta{\textrm{Im}}\bar{\cal{S}}^{\textrm{out}}_{1}(r,t)}{\delta k}=\frac{{\textrm{Im}}{\cal{S}}^{\textrm{out}}_{1}(r,t)-{\textrm{Im}}\bar{\cal{S}}^{\textrm{out}}_{1}(r,t)}{\delta k}\nonumber\\
=-\frac{8\Omega\beta_1}{(D-2){r_\textrm{h}}^{(D-2)}}{\textrm{Im}}\int_C{\frac{dr}{g(r)}}.
\label{transformed2}
\end{eqnarray}
Considering the scalar field action (\ref{action3}) it can be shown that under a constant scale transformation ${\bar g}_{\mu\nu}$, the action is not invariant in the presence of trace anomaly, since  
\begin{eqnarray}
\frac{\delta {{{\textrm{Im}}\cal S}}}{\delta k}=\frac{1}{2}\textrm{Im}\int{d^4x\sqrt{-g}(\langle T^{\mu}_{~\mu} \rangle^{(1)}+\langle T^{\mu}_{~\mu} \rangle^{(2)}+...)},
\label{trace}
\end{eqnarray} 
 where $\langle T^{\mu}_{~\mu} \rangle^{(i)}$' s  are the trace of the regularized stress energy tensor calculated for $i$-th loop. Considering one loop only and comparing (\ref{transformed2}) and (\ref{trace}) we find the coefficient of the leading correction to the semiclassical entropy as
\begin{eqnarray}
\beta_1 &=& -\left(\textrm{Im}\int_C\frac{dr}{g(r)}\right)^{-1}\frac{(D-2)r^{(D-2)}_{\textrm{h}}}{16\Omega}{\textrm{Im}\int{d^{D}x\sqrt{-g}\langle T^{\mu}~_{\mu} \rangle^{(1)}}}\nonumber\\
        &=& -\frac{(D-2)r^{(D-2)}_h g'(r_h)}{16\pi\Omega}{\textrm{Im}\int{d^{D}x\sqrt{-g}\langle T^{\mu}~_{\mu} \rangle^{(1)}}}.
\label{beta1}
\end{eqnarray}
Note that this is the undetermined coefficient ($\beta_1$) that occurs in the expressions of the corrected entropy (\ref{3.2}), (\ref{3.4}) and (\ref{3.5}). 

For Gauss-Bonnet dS black holes we have found the expression of the corrected entropy associated with the cosmological horizon (\ref{3.6}). We can follow a strategy, similar to the previous one, to relate $\beta_1$ appearing in (\ref{3.6}), with the trace anomaly. For that first note that in the case of cosmological horizon the ingoing mode can always cross the cosmological horizon to reach the observer. Therefore, unlike the black hole horizon, here one has $P_{in}=1$ in the classical limit and from (\ref{trace2}) it follows that
\begin{eqnarray}
{\textrm{Im}~t} &=& {\textrm{Im}}\int_C\frac{dr}{g(r)}
\label{trace4}
\end{eqnarray} 
and consequently
\begin{eqnarray}
{\textrm{Im}}{\cal{S}}^{\textrm{in}}_{0}(r,t) &=& 2\Omega{\textrm{Im}}\int_C{\frac{dr}{g(r)}}\label{ims02}.
\label{trace5}
\end{eqnarray}  
Now using the infinitesimal scale transformation to the metric coefficients and following the earlier methodology one would find $\beta_1$ appearing in (\ref{3.6}) as
\begin{eqnarray}
\beta_1|_{\textrm{ch}} &=& \left(\textrm{Im}\int_C\frac{dr}{g(r)}\right)^{-1}\frac{(D-2)r_c^{(D-2)}}{16\Omega}{\textrm{Im}\int{d^{D}x\sqrt{-g}\langle T^{\mu}~_{\mu} \rangle^{(1)}}}\nonumber\\
                       &=& -\frac{(D-2)r_c^{(D-2)}g'(r_c)}{16\pi\Omega}{\textrm{Im}\int{d^{D}x\sqrt{-g}\langle T^{\mu}~_{\mu} \rangle^{(1)}}}
\label{cosbeta1}
\end{eqnarray}
Here the contour is chosen as a semicircle in the lower region of the complex plane, going from right to left, starting from just outside the cosmological horizon to the inner region avoiding the singularity at the cosmological horizon. 

It is evident from (\ref{beta1}) and (\ref{cosbeta1}) that the explicit calculation of $\beta_1$, therefore, needs prior knowledge of the trace anomaly in `$D$' dimensions. In the following we shall take specific metrics to give a simplified expression for $\beta_1$. However, since there is no known result of trace anomaly for any dimension $D>4$ we shall not be able to give an explicit value for $\beta_1$.

\subsection{Spherically symmetric, asymptotically flat black holes}
We can simplify $\beta_1$ in (\ref{beta1}) by using $g(r)=f(r)$ from (\ref{1.6}), (\ref{1.6a}), (\ref{1.7a}) to give the corrected entropy (\ref{3.2}). Also since $\Omega$ in (\ref{actionform}) is the conserved quantity corresponding to the time translation Killing vector field, it is nothing but the total gravitational energy or equivalently the mass of the black hole, given by (\ref{1.7a}). Using these we get 
\begin{eqnarray}
\beta_1^{\textrm{flat}}=-\frac{2G[(D-3)r_h^2+\frac{\bar\lambda}{2}(D-5)]r_h^2}{(r_h^2+\bar\lambda)(2r_h^2+\bar\lambda)A_{D-2}}{\textrm{Im}\int{d^{D}x\sqrt{-g}\langle T^{\mu}~_{\mu} \rangle^{(1)}}}.
\label{beta1flat}
\end{eqnarray}
The expression of trace anomaly in the background of a curved spacetime is not known for dimensions greater than four. Therefore we cannot integrate this expression in general to find $\beta_1$. Now the corrected entropy of the spherically symmetric asymptotically flat black holes can be found from (\ref{3.2}) and (\ref{beta1flat}) as
\begin{eqnarray}
S_{\textrm{bh}}^{\textrm{flat}}=\frac{A_{D-2}r_h^{(D-2)}}{4G\hbar}\left(1+\frac{D-2}{D-4}\frac{\bar\lambda}{r_h^2}\right)-\frac{[(D-3)r_h^2+\frac{\bar\lambda}{2}(D-5)]r_h^2}{2(r_h^2+\bar\lambda)(2r_h^2+\bar\lambda)}\times\nonumber\\
\left(\int{d^{D}x\sqrt{-g}\langle T^{\mu}~_{\mu} \rangle^{(1)}}\right)\left(\log(\frac{r_h^{D-2}}{G\hbar}) -\frac{(D-2)\bar\lambda}{2r_h^2}\right)+{\cal O}(\hbar). 
\label{corenflat}
\end{eqnarray}  
For $D=4$ the expression for the trace anomaly is known \cite{Dewitt} and the integration can be performed. We did this in our previous works for the Schwarzschild black hole  \cite{Modak,Majhitrace}. The coefficient of the leading correction to entropy was found to be $\frac{1}{90}$ for that case.

\subsection{Topological Gauss-Bonnet AdS black holes}
The metric coefficients $g(r)=F(r)$ are given by (\ref{1.11}). Substituting this in (\ref{beta1}) one can perform the contour integration around the event horizon ($r_h$) and replacing $\Omega$ by the total gravitational mass ($M$) we find the coefficient of the leading non-logarithmic correction to the entropy as
\begin{eqnarray}
\beta_1^{\textrm{AdS}}=-\frac{G[(D-1)r_h^4+(D-3)kl^2r_h^2+(D-5)\tilde\alpha k^2l^2]}{(r_h^2+2\tilde\alpha k)(k+\frac{\tilde\alpha k^2}{r_h^2}+\frac{r_h^2}{l^2})\Sigma_k l^2}{\textrm{Im}\int{d^{D}x\sqrt{-g}\langle T^{\mu}~_{\mu} \rangle^{(1)}}}.
\label{beta1ads}
\end{eqnarray}
Therefore the expression of the corrected entropy is given by
\begin{eqnarray}
S_{\textrm{bh}}^{\textrm{AdS}} &=& \frac{\Sigma_kr_h^{(D-2)}}{4G\hbar}\left(1+\frac{D-2}{D-4}\frac{2\tilde\alpha k}{r_h^2}\right)-\frac{(D-1)r_h^4+(D-3)kl^2r_h^2+(D-5)\tilde\alpha k^2l^2}{4l^2(r_h^2+2\tilde\alpha k)(k+\frac{\tilde\alpha k^2}{r_h^2}+\frac{r_h^2}{l^2})}\times\nonumber\\
 && \left(\int{d^{D}x\sqrt{-g}\langle T^{\mu}~_{\mu} \rangle^{(1)}}\right)\left(\log(\frac{r_h^{D-2}}{G\hbar}) -\frac{(D-2)\tilde\alpha k}{r_h^2}\right)+{\cal O}(\hbar). 
\label{corenads}
\end{eqnarray}

\subsection{Gauss-Bonnet dS black holes}
Let us first consider the corrected entropy associated with the black hole horizon (\ref{3.5}). Performing the contour integral in (\ref{beta1}) around the black hole horizon and identifying $\Omega$ as the gravitational AD mass (\ref{1.17a}), we get 
\begin{eqnarray}
\beta_1^{\textrm{dS}}\big|_{\textrm{bh}}=-\frac{G[(D-3)l^2r_h^2-(D-1)r_h^4+(D-5)\tilde\alpha l^2]}{A_{D-2}(r_h^2+2\tilde\alpha)(1+\frac{\tilde\alpha}{r_h^2}-\frac{r_h^2}{l^2}) l^2}{\textrm{Im}\int{d^{D}x\sqrt{-g}\langle T^{\mu}~_{\mu} \rangle^{(1)}}}.
\label{beta1ds1}
\end{eqnarray}
Therefore the corrected entropy associated with the black hole horizon for the dS black holes is given by 
\begin{eqnarray}
S_{\textrm{bh}}^{\textrm{dS}} &=& \frac{A_{D-2}r_h^{(D-2)}}{4G\hbar}\left(1+\frac{D-2}{D-4}\frac{2\tilde\alpha }{r_h^2}\right)-\frac{(D-3)r_h^2l^2-(D-1)r_h^4+(D-5)\tilde\alpha l^2}{4l^2(r_h^2+2\tilde\alpha)(1+\frac{\tilde\alpha}{r_h^2}-\frac{r_h^2}{l^2})}\times\nonumber\\
&& \left(\int{d^{D}x\sqrt{-g}\langle T^{\mu}~_{\mu} \rangle^{(1)}}\right)\left(\log(\frac{r_h^{D-2}}{G\hbar}) -\frac{(D-2)\tilde\alpha }{r_h^2}\right)+{\cal{O}}(\hbar). 
\label{corends1}
\end{eqnarray} 

For the cosmological horizon the identification of $\Omega$ is done with the BBM mass which is just the negative of the AD mass with $r_c$ replacing $r_h$. The coefficient of the leading correction is then simplified as,
\begin{eqnarray}
\beta_1^{\textrm{dS}}\big|_{\textrm{ch}}=-\frac{G[(D-3)l^2r_c^2-(D-1)r_c^4+(D-5)\tilde\alpha l^2]}{A_{D-2}(r_c^2+2\tilde\alpha)(1+\frac{\tilde\alpha k^2}{r_c^2}-\frac{r_c^2}{l^2})l^2}{\textrm{Im}\int{d^{D}x\sqrt{-g}\langle T^{\mu}~_{\mu} \rangle^{(1)}}}.
\label{beat1ds2}
\end{eqnarray} 
The corrected entropy of cosmological horizon of the Gauss-Bonnet dS black hole now follows from (\ref{3.6}) and (\ref{beat1ds2}), 
\begin{eqnarray}
S_{\textrm{ch}}^{\textrm{dS}} &=& \frac{A_{D-2}r_c^{(D-2)}}{4G\hbar}\left(1+\frac{D-2}{D-4}\frac{2\tilde\alpha }{r_c^2}\right)+\frac{(D-5)\tilde\alpha l^2-(D-1)r_c^4+(D-3)l^2r_c^2}{4l^2(r_c^2+2\tilde\alpha)(1+\frac{\tilde\alpha}{r_c^2}+\frac{r_c^2}{l^2})}\times\nonumber\\
 && \left(\int{d^{D}x\sqrt{-g}\langle T^{\mu}~_{\mu} \rangle^{(1)}}\right)\left(\log(\frac{r_c^{D-2}}{G\hbar}) -\frac{(D-2)\tilde\alpha}{r_c^2}\right)+{\cal O}(\hbar).
\label{corends2}
\end{eqnarray}


\section{Conclusions}{\label{sec:7}}
Let us now summarize the work carried out in this paper. We showed, adopting the tunneling method advocated in \cite{Majhiflux,Majhiconnect}, that Lovelock black holes do emit scalar particles and fermions with a perfect blackbody spectrum with a temperature given by the semiclassical Hawking temperature. This result, which is a new finding, was derived for both black hole (event) horizon and cosmological horizon of Lovelock black holes. It was also found that in the presence of quantum corrections to the WKB ansatz the blackbody nature of the modified radiation spectrum does not change. Greybody factors were absent, rather the temperature received some corrections. We calculated the corrected Hawking temperature for both the black hole (event) horizon and also for the cosmological horizon for different spacetimes in Lovelock gravity.

The temperature corresponding to the modified spectrum, as calculated by us, reproduced the Hawking temperature at the lowest order. Using the first law of thermodynamics the modified entropy corresponding to the corrected Hawking temperatures was calculated for different cases. It is a well known fact that Lovelock black holes do not obey the semiclassical area law \cite{Myers,Cai,Cai2}. This modified semiclassical area law was reproduced here as the lowest order contribution. One of the new results reported in this paper was that the higher order corrections to entropy did not involve the famous logarithmic and inverse area corrections. The modified corrections were shown to be a consequence of the Gauss-Bonnet term. When the coupling constant of the Gauss-Bonnet term vanished one was left with the semiclassical area law. Also, the standard logarithmic and inverse area corrections were reproduced in the higher orders. We expressed the coefficient ($\beta_1$) of the leading (non logarithmic) correction in terms of the trace anomaly of the stress tensor by making infinitesimal scale transformation of the metric coefficients.

Some useful definitions were summarised in the appendices. Particularly, Kruskal-like extensions for the black hole (event) and cosmological horizons were performed here. \\\\\\   

\section{Acknowledgments}

One of the authors (S.K.M) thanks the Council of Scientific and Industrial Research (C.S.I.R), Government of India, for financial support.\\\\\\ 

\appendix

\section{Kruskal-like extension for the black hole (event) horizon}\label{sec:A}

To perform a Kruskal-like extension of a general chargeless, static metric (\ref{metric}), we first define the tortoise coordinate as
\begin{eqnarray}
r^*=\int{\frac{dr}{g(r)}}.
\label{tortoise}
\end{eqnarray}
It is known that the spacetime structure of the extended regions are extremely sensitive to the different choices for $g(r)$. In order to study the behaviors of the outgoing and ingoing modes with respect to the black hole event horizon ($r_{h}$) we actually need to see only the behavior of the spacetime in a very narrow region just inside and outside of $r_h$. Therefore, for our purpose, we first take the near horizon limit of the metric coefficient,
\begin{eqnarray}
g(r)=(r-r_h)g'(r_h)+\frac{(r-r_h)^2}{2}g''(r_h)+{\cal O}(r-r_h)^3
\label{nearhor}
\end{eqnarray}

In the following we shall consider two different cases for inside and outside (where an observer is present) the event horizon to show that the same spacetime metric is valid in both the regions.

{\bf Case I}: When $r=r_{\textrm{obs}}>r_h$ (Outside the event horizon): At a distance ($\rho$) just outside the horizon, $r=r_h+\rho$, such that $|\rho|<<r_h$, one has  $dr=d\rho$ and
\begin{eqnarray}
g(r)=\rho g'(r_h)+\frac{\rho^2}{2}g''(r_h)+{\cal O}(\rho^3)
\label{nearhor2}
\end{eqnarray}
Therefore (\ref{tortoise}) can be integrated over this narrow region to yield
\begin{eqnarray}
r^*_{obs} &=& \frac{1}{g'(r_h)}\ln[\frac{\rho}{g'(r_h)+\rho g''(r_h)/2}]\nonumber\\
                   &=& \frac{1}{g'(r_h)}\ln[\frac{(r-r_h)}{g'(r_h)+(r-r_h) g''(r_h)/2}]
\label{tortoise2}
\end{eqnarray}
The advanced and retarded time coordinates, in this region, are usually defined by 
\begin{eqnarray}
v_{obs} &=& t_{obs}+r^*_{obs}\nonumber\\
u_{obs} &=& t_{obs}-r^*_{obs}
\label{null}
\end{eqnarray}
respectively. With these definitions (\ref{metric}) becomes
\begin{eqnarray}
ds^2=-g(r)du_{obs}dv_{obs}+r^2d\Omega_{D-2}^2
\label{nullmetric}
\end{eqnarray}
Now we make the following two successive coordinate transformations
\begin{eqnarray}
V_{obs} &=& e^{\kappa v_{obs}}\nonumber\\
U_{obs} &=& -e^{-\kappa u_{obs}}.
\label{capital}
\end{eqnarray}
Also, 
\begin{eqnarray}
T_{obs} &=& \frac{(V_{obs}+U_{obs})}{2}\nonumber\\
X_{obs} &=& \frac{(V_{obs}-U_{obs})}{2},
\label{capital2}
\end{eqnarray}
 where $\kappa$ is a constant that will be identified later with the surface gravity. With these transformations we find the desired form of the $r-t$ sector of the metric (\ref{metric}) in Kruskal coordinates,
\begin{eqnarray}
ds^2=-\frac{g(r)}{\kappa^2}e^{-2\kappa r^*_{obs}}(dT_{obs}^2-dX_{obs}^2).
\label{kruskal}
\end{eqnarray}
Putting the values of $g(r)$ from (\ref{nearhor}) and $r^*$ from (\ref{tortoise2}) into the spacetime interval (\ref{kruskal}) and simplifying the pre-factor by choosing $\kappa$ to be the surface gravity ($\kappa=g'(r_\textrm{h})/2$), we find
\begin{eqnarray}
ds^2=\frac{4\left(g'(r_h)^2+\frac{3g'(r_h)g''(r_h)}{2}(r-r_h)+\frac{g''(r_h)^2}{2}(r-r_h)^2\right)}{g'(r_h)^2}(-dT_{obs}^2+dX_{obs}^2).
\label{kruskal2}
\end{eqnarray}
Now using (\ref{capital}) and (\ref{capital2}) we find the Kruskal-like coordinates which are valid outside the event horizon where the observer is present, are given by 
{\begin{eqnarray}
T_{obs} &=& e^{\kappa r_{{obs}}^{*}}\sinh{\kappa t_{obs}}\nonumber\\
X_{obs} &=& e^{\kappa r^*_{obs}}\cosh{\kappa t_{obs}}.
\label{transformation}
\end{eqnarray}
This clearly shows, irrespective of choosing any particular $g(r)$, at the event horizon ($r=r_h$ or $\rho=0$) one does not have any spacetime singularity in this Kruskal-like extension (\ref{kruskal2}). Therefore one can easily make the extension of the spacetimes (\ref{1.5}), (\ref{1.10}) and (\ref{1.15}) with appropriate choice of coordinates. The same is true if one considers higher order terms in the expansion (\ref{nearhor}). The only finite contribution to the interval (\ref{kruskal2}) comes from the linear term in the expansion (\ref{nearhor}) while others vanish at $r=r_h$. This consistency is essential for the calculation of the blackbody spectrum.

{\bf Case II}: When $r=r_{\textrm{in}}<r_h$ (Inside the event horizon): Let us now consider a situation at a distance $\rho$ inside the event horizon. Here $r=r_{\textrm{in}}=r_h-\rho$ and $dr=-d\rho$. Using the same expansion (\ref{nearhor2}) and integrating (\ref{tortoise}), we get the required expression for the tortoise coordinate inside the event horizon, 
\begin{eqnarray}
r^*_{in} &=& \frac{1}{g'(r_h)}\ln[\frac{\rho}{g'(r_h)-\rho g''(r_h)/2}]\nonumber\\
                   &=& \frac{1}{g'(r_h)}\ln[\frac{(r_h-r)}{g'(r_h)-(r_h-r) g''(r_h)/2}].
\label{tortoise3}
\end{eqnarray}
Consider the following coordinate transformations,
\begin{eqnarray}
v_{in} &=& t_{in}+r^*_{in}\nonumber\\
u_{in} &=& t_{in}-r^*_{in},
\label{null2}
\end{eqnarray} 
and
\begin{eqnarray}
V_{in} &=& e^{\kappa v_{in}}\nonumber\\
U_{in} &=& e^{-\kappa u_{in}}.
\label{capital3}
\end{eqnarray}
Also
\begin{eqnarray}
T_{in} &=& \frac{(V_{in}+U_{in})}{2}\nonumber\\
X_{\textrm{in}} &=& \frac{(V_{in}-U_{in})}{2}.
\label{capital4}
\end{eqnarray}
With these coordinate transformations in (\ref{metric}) we are finally left with a metric whose $r-t$ sector is given by 
\begin{eqnarray}
ds^2=\frac{4\left(g'(r_h)^2+\frac{3g'(r_h)g''(r_h)}{2}(r-r_h)+\frac{g''(r_h)^2}{2}(r-r_h)^2\right)}{g'(r_h)^2}(-dT_{in}^2+dX_{{in}}^2).
\label{kruskal3}
\end{eqnarray}
The new Kruskal coordinates which are valid inside the event horizon are now found from (\ref{null2}), (\ref{capital3}) and (\ref{capital4}),
{\begin{eqnarray}
T_{in} &=& e^{\kappa r^*_{in}}\cosh{\kappa t_{in}}\nonumber\\
X_{in} &=& e^{\kappa r^*_{in}}\sinh{\kappa t_{in}}.
\label{transformation2}
\end{eqnarray}
One can now realize that all the coordinate transformations used here are identical to the previous case for $r>r_{\textrm h}$ except for the definition of $U_{in}$. There is a relative sign difference between the functional choice of $U_{in}$ and  $U_{obs}$. This new definition ensures that the inner portion of the extended spacetime metric remains timelike. If one follows the earlier definition he/she will end up with a spacelike metric interval which we want to avoid because there is no coordinate singularity at the event horizon. Therefore one can conclude that  
\begin{eqnarray}
ds^2=\frac{4\left(g'(r_h)^2+\frac{3g'(r_h)g''(r_h)}{2}(r-r_h)+\frac{g''(r_h)^2}{2}(r-r_h)^2\right)}{g'(r_h)^2}(-dT^2+dX^2)
\label{kruskal4}
\end{eqnarray} 
is the only metric which is valid in a narrow region on both sides of the event horizon and the Kruskal coordinates outside (where the observer is present) and inside the event horizon are defined by (\ref{transformation}) and (\ref{transformation2}) respectively. 

 
\section{Kruskal-like extension for the cosmological horizon}\label {sec:B}
To see the behaviour of the spacetime (\ref{metric}) at the cosmological horizon it is required to expand $g(r)$ near the cosmological horizon ($r_h$), given by 
\begin{eqnarray}
g(r)=(r-r_c) g'(r_c)+\frac{(r-r_c)}{2}g''(r_c)+{\cal O}((r-r_{c})^3).
\label{nearhor3}
\end{eqnarray}
The tortoise coordinates inside (where the observer is present; $r=r_{{obs}}<r_{c}$) and outside ($r=r_{{out}}>r_{c}$) the cosmological horizon are defined by
\begin{eqnarray}
r^*_{obs}= \frac{1}{g'(r_c)}\ln[\frac{(r_c-r)}{g'(r_c)-(r_c-r) g''(r_c)/2}]
\label{torcos2}
\end{eqnarray}
and 
\begin{eqnarray}
r^*_{out}= \frac{1}{g'(r_c)}\ln[\frac{(r-r_c)}{g'(r_c)+(r-r_c) g''(r_c)/2}]
\label{torcos1}
\end{eqnarray}
respectively. The sets of null coordinates inside and outside the cosmological horizon are defined as
\begin{eqnarray}
v_{obs} &=& t_{obs}+r^*_{obs}\nonumber\\
u_{obs} &=& t_{obs}-r^*_{obs}
\label{cosnull}
\end{eqnarray} 
and 
\begin{eqnarray}
v_{out} &=& t_{out}+r^*_{out}\nonumber\\
u_{out} &=& t_{out}-r^*_{out}
\label{cosnull2}
\end{eqnarray} 
respectively. Now by exactly mimicking the methodology developed for the case of black hole horizon it can be shown that the metric 
\begin{eqnarray}
ds^2=\frac{4\left(g'(r_c)^2+\frac{3g'(r_{c})g''(r_{c})}{2}(r-r_c)+\frac{g''(r_{c})^2}{2}(r-r_c)^2\right)}{g'(r_c)^2}(-dT^2+dX^2)
\label{kruskal5}
\end{eqnarray}
is defined both inside and outside the horizon. This is the analogue of (\ref{kruskal4}). This time the Kruskal-like coordinates in the region inside and outside the cosmological horizon are, respectively, given by
\begin{eqnarray}
T_{obs} &=& e^{\kappa r^*_{obs}}\cosh{\kappa t_{obs}}\nonumber\\
X_{obs} &=& e^{\kappa r^*_{obs}}\sinh{\kappa t_{obs}}
\label{transformation4}
\end{eqnarray}
and
\begin{eqnarray}
T_{out} &=& e^{\kappa r^*_{out}}\sinh{\kappa t_{out}}\nonumber\\
X_{out} &=& e^{\kappa r^*_{out}}\sinh{\kappa t_{out}},
\label{transformation3}
\end{eqnarray}
where $\kappa=\frac{g'(r_c)}{2}$ is the surface gravity at the cosmological horizon.


\section{Identification of right and left moving modes}\label {sec:C}
To identify different modes in different regions we use the following convention. If the eigenvalue of the radial momentum operator $\hat p(r)$ , while acting on a specific solution of the semiclassical (WKB) mode (\ref{soln}), is positive, then the mode is right-moving (outgoing). Similarly a left-moving (incoming) mode corresponds to a negative eigenvalue. In this convention the spacelike nature of $\hat p(r)$ must be kept unchanged in any region (i.e. inside or outside the horizon) of the spacetime. 
 
{\bf Black hole horizon:} For all Lovelock black holes, considered in this paper, $\Omega$ (which is the conserved quantity to the timelike Killing vector) is nothing but the mass ($M$) of the black hole. In a region outside the event horizon (position of the observer) ($r>r_h$), $\hat p(r)=-i\hbar{\frac{\partial}{\partial r}}$ and the left ($L$) or right ($R$) moving modes are found as
\begin{eqnarray}
\Phi^{L}_{obs} &=& e^{-\frac{i}{\hbar}\Omega v_{obs}}\label{conv1}\\
\Phi^{R}_{obs} &=& e^{-\frac{i}{\hbar}\Omega u_{obs}}\label{conv2}.
\end{eqnarray}       
To keep the spacelike nature of the momentum operator inside the black hole event horizon, one must use the definition $\hat p(r)=i\hbar{\frac{\partial}{\partial r}}$. Using this one can identify
\begin{eqnarray}
\Phi^{L}_{in} &=& e^{-\frac{i}{\hbar}\Omega v_{in}}\label{conv3}\\
\Phi^{R}_{in} &=& e^{-\frac{i}{\hbar}\Omega u_{in}}\label{conv4}.
\end{eqnarray}
as the left and right moving modes respectively.

{\bf Cosmological horizon:} For the Gauss-Bonnet dS black hole we have an extra cosmological horizon. Here $\Omega$ or equivalently the gravitational mass is given by the BBM mass \cite{BBM} which is negative. In a region inside the cosmological horizon where the observer is present ($r_h<r<r_c$), $\hat p(r)=-i\hbar\frac{\partial}{\partial r}$, and one can identify the left and right moving modes as
\begin{eqnarray}
\Phi^{L}_{obs} &=& e^{-\frac{i}{\hbar}\Omega u_{obs}}\label{conv5}\\
\Phi^{R}_{obs} &=& e^{-\frac{i}{\hbar}\Omega v_{obs}}\label{conv6}.
\end{eqnarray}  
Similarly outside the cosmological horizon ($r>r_c$), one has $\hat p(r)=i\hbar\frac{\partial}{\partial r}$, therefore one finds 
\begin{eqnarray}
\Phi^{L}_{out} &=& e^{-\frac{i}{\hbar}\Omega u_{out}}\label{conv7}\\
\Phi^{R}_{out} &=& e^{-\frac{i}{\hbar}\Omega v_{out}}\label{conv8},
\end{eqnarray}  
as the left and right moving modes respectively.


\end{document}